# Possible thermochemical disequilibrium in the atmosphere of the exoplanet GJ 436b

Kevin B. Stevenson[1], Joseph Harrington[1], Sarah Nymeyer[1], Nikku Madhusudhan[2], Sara Seager[2], William C. Bowman[1], Ryan A. Hardy[1], Drake Deming[3], Emily Rauscher[4] & Nate B. Lust[1]

[1]Planetary Sciences Group, Department of Physics, University of Central Florida, Orlando, Florida 32816, USA. [2]Department of Physics and Department of Earth, Atmospheric and Planetary Sciences, Massachusetts Institute of Technology, Cambridge, Massachusetts 02159, USA. [3]Planetary Systems Laboratory, NASA Goddard Space Flight Center, Greenbelt, Maryland 20771, USA. [4]Department of Astronomy, Columbia University, New York, New York 10027, USA.

**The nearby extrasolar planet GJ 436b—which has been labelled as a 'hot Neptune'—reveals itself by the dimming of light as it crosses in front of and behind its parent star as seen from Earth. Respectively known as the primary transit and secondary eclipse, the former constrains the planet's radius and mass[1,2], and the latter constrains the planet's temperature[3,4] and, with measurements at multiple wavelengths, its atmospheric composition. Previous work[5] using transmission spectroscopy failed to detect the 1.4-μm water vapour band, leaving the planet's atmospheric composition poorly constrained. Here we report the detection of planetary thermal emission from the dayside of GJ 436b at multiple infrared wavelengths during the secondary eclipse. The best-fit compositional models contain a high CO abundance and a substantial methane ($CH_4$) deficiency relative to thermochemical equilibrium models[6] for the predicted hydrogen-dominated atmosphere[7,8]. Moreover, we report the presence of some $H_2O$ and traces of $CO_2$. Because $CH_4$ is expected to be the dominant carbon-bearing species, disequilibrium processes such as vertical mixing[9] and polymerization of methane[10] into substances such as ethylene may be required to explain the hot Neptune's small $CH_4$-to-CO ratio, which is at least $10^5$ times smaller than predicted[6].**

Using the Spitzer Space Telescope[11], the Spitzer Exoplanet Target of Opportunity program observed multiple secondary eclipses at wavelengths of 3.6, 4.5, 5.8, 8.0, 16 and 24 μm. Previous analyses[3,4] of our 8.0-μm secondary eclipse data confirm an eccentric orbit around GJ 436, which is a cool, M-dwarf star. Standard image calibration and photometry produced light curves (tables of system flux versus time at each wavelength) that are available as Supplementary Information, as are details of centring and photometry. Some channels have well documented systematic effects that our Metropolis-Hastings Markov-chain Monte Carlo (MCMC) model[12] fits simultaneously with the eclipse parameters. Systematics include positional sensitivity variation[13] at 3.6, 4.5 and 5.8 μm, where the measured flux correlates with the sub-pixel location of the stellar centre, and time-varying sensitivity[14] at 4.5, 5.8, 8.0 and 16 μm. Responsivity of the 24-μm channel is relatively stable[15]. Figure 1 shows the observed secondary eclipses with best-fit models, and Table 1 presents the relevant eclipse parameters.

The phase of secondary eclipse imposes a tight constraint on the planet's eccentricity, $e$, and argument of periapsis, $\omega$. Using the secondary eclipse times listed in Table 1, in addition to published transit[16] and radial-velocity data[17], a single-planet Keplerian orbit for GJ 436b has a period of 2.6438983 ± 0.0000016 days and an ephemeris time of Julian date 2,454,222.61587 ± 0.00012 (all errors are $1\sigma$). These are nearly identical to the published results[16], which do not consider secondary eclipses. Using either result, the weighted average of the five measured secondary eclipse phases is 0.5868 ± 0.0003. This significant improvement from previous analyses[3,4] is due to the more precise ephemeris time and the use of multiple secondary eclipses over a long baseline. The weighted average of the minimum eccentricities, defined as $e_{\min} \approx e \cos(\omega)$, is 0.1368 ± 0.0004. Using $\omega = 351 \pm 1.2°$ (ref. 17), we find $e = 0.1385 \pm 0.0006$. To compute all of the orbital parameters (Supplementary Information), we used the published results referenced above in addition to the eclipse times presented here. Our best-fit value for $e$ is $0.1371^{+0.0048}_{-0.00013}$.

Our broadband observations constrain a one-dimensional atmospheric model, using a new temperature and abundance retrieval method[18]. This method searches over a wide parameter space using a functional form for the pressure–temperature profile (based on prior 'hot Jupiter' and Solar System studies), a grid of abundance combinations, and energy conservation. We calculated ~$10^6$ models, which considered both inversion and non-inversion temperature profiles and abundances that varied over several orders of magnitude per constituent. Figure 2 shows two representative models (the red and blue lines) that fit the data, and Table 2 compares them to seven other objects with hydrogen-dominated atmospheres. The red model has a dayside-to-nightside energy redistribution ratio of <0.04; the blue model favours a more efficient distribution ratio of <0.31. The red model fits the data better.

Chemical equilibrium predicts $H_2$, $H_2O$, $CH_4$, CO and $NH_3$ to be the most abundant molecules in GJ 436b's atmosphere (helium must also be present but contributes minimally to the spectrum and to active chemistry). Conventional chemical composition models predict[6] the major emission contributions to come from spectroscopically active $H_2O$, $CH_4$, and, to a lesser extent, CO, and possibly $CO_2$. In a reduced, hydrogen-dominated atmosphere at ~700 K, $CH_4$ is thermochemically favoured to be the main carbon-bearing molecule. Assuming solar abundances for the elements and the pressure–temperature profile shown in Supplementary Fig. 5, chemical equilibrium predicts[6] a $CH_4$-to-$H_2$ mixing ratio of $7 \times 10^{-4}$ and an $H_2O$ mixing ratio of $2 \times 10^{-3}$. However, the strong planetary emission at 3.6 μm, combined with the non-detection at 4.5 μm, calls for a methane abundance that is depleted by a factor of ~7,000. The low $H_2O$ abundance favoured by our red model could, in principle, result from carbon and oxygen abundances that are ~0.01 times solar values; however, the resulting $CH_4$ mixing ratio would still be too high, by two orders of magnitude, to explain the data.

Methane absorbs strongly in the 3.6-μm band. CO and $CO_2$ have absorption features at 4.5 μm, $CO_2$ being the stronger absorber. The high flux at 3.6 μm suggests very low absorption due to methane, while the low flux at 4.5 μm implies high absorption due to CO and/or $CO_2$. The degeneracy between the two molecules is solved by the low $CO_2$ concentration needed at 16 μm. The absence of observed flux in the 4.5-μm channel thus requires large amounts of CO, which is not expected in such a reduced atmosphere under thermochemical equilibrium, and makes a future detection at 4.5 μm important.

The flux modulation at 3.6 μm is our strongest detection, with a signal-to-noise ratio of 12.1, and has been confirmed by an independent analysis. Using $2\sigma$ error bars for this observation and the $3\sigma$ upper limit at 4.5 μm, the low-methane requirement cannot be lifted (this result is relatively insensitive to the remaining wavelengths). An increased methane mixing ratio of $10^{-6}$ would result in a higher blackbody continuum, thus requiring a $CO_2$ mixing ratio $\geq 10^{-3}$ in order to fit the flux constraint at 4.5 μm (Figure 3). However, thermochemical equilibrium and photochemical models predict a $CO_2$ mixing ratio of $\sim 10^{-7}$ in hydrogen-dominated atmospheres at solar abundance ($\sim 10^{-5}$ for 30 times solar metallicities)[19,20].

We also explored other possibilities to explain the observations. A temperature inversion does not fit the data well, assuming thermochemical equilibrium, because $H_2O$ and $CH_4$ would emit much more strongly than we observe in the 5.8- and 8.0-μm channels, respectively. Non-local thermodynamic equilibrium emission from the dayside of exoplanet HD 189733b is attributed[21] to $CH_4$ fluorescence near 3.25 μm. However, our 3.6-μm detection is too strong to be explained by fluorescence alone. Alternatively, the methane deficiency could be explained by a lack of hydrogen; however, mass and radius constraints placed by transit and radial-velocity observations call for a hydrogen-dominated atmosphere[7,8], which we explored above. Atmospheric compositions dominated by an alternative species (such as He or $N_2$) are difficult to invoke plausibly. Hydrogen is the most abundant species in planet-forming disks and atmospheric escape rates are small for Neptune-mass planets. Although the observations were not made simultaneously, planet variability and stellar activity are unlikely explanations for our observations. A global, planetary temperature variation of 400 K manifesting in 2.64 days (the time between the 3.6- and 4.5-μm observations) would be unprecedented in planetary science, as would a transient hot vortex[22] with one-third the planetary radius and $T \approx 2{,}200$ K that appeared during only one of our six observations. Stellar activity, which is common among M dwarfs, would need to be timed precisely with the secondary eclipse for us not to detect and mask it.

The brown dwarf GJ 570D has an effective temperature similar to that of GJ 436b (800 K), but at atmospheric levels where $T < 1{,}100$ K, $CH_4$ is the dominant carbon-bearing molecule, with a $CH_4$-to-CO ratio of $\sim 10^2$ (ref. 9). We estimate a ratio of $\leq 1 \times 10^{-3}$ for GJ 436b (Table 2); however, the exoplanet is strongly irradiated on one side, which can drive atmospheric dynamics and disequilibrium chemistry. Vertical mixing[9] can dredge CO up from deeper and hotter parts of the atmosphere, where CO is favoured, resulting in a small $CH_4$-to-CO ratio if the rate of dredging is faster than the rate of the reaction that converts CO to $CH_4$ ($CO + 3H_2 \leftrightarrow CH_4 + H_2O$). However, the observed $CH_4$-to-CO mixing ratio would require large amounts of vertical mixing. Alternatively, $CH_4$ may be depleted by polymerization into hydrocarbons such as ethylene ($C_2H_4$). This is a major methane reaction pathway at these temperatures[10]. These possibilities represent starting points for future theoretical work with this atmosphere.

**Supplementary Information** is linked to the online version of the paper at www.nature.com/nature.

**Acknowledgements** We thank the Spitzer staff for rapid scheduling; M. Gillon, A. Lanotte and T. Loredo for discussions; D. Wilson for contributed code; and A. Wright for manuscript comments. We thank the following for software: the Free Software Foundation, W. Landsman and other contributors to the Interactive Data Language Astronomy Library, contributors to SciPy, Matplotlib, and the Python programming language, and the open-source community. This work is based on observations made with the Spitzer Space Telescope, which is operated by the Jet Propulsion Laboratory, California Institute of Technology, under a contract with NASA. This material is based on work supported by the US NSF and by the US NASA through an award issued by JPL/Caltech.

**Author Contributions** K.B.S. wrote the paper and Supplementary Information with contributions from J.H., N.M. and R.A.H.; N.M. and S.S. produced the atmospheric models; S.N., K.B.S. and W.C.B. reduced the data; K.B.S., J.H. and D.D. analysed the results; D.D. ran an independent analysis; R.A.H. produced the orbital parameter results; and J.H., K.B.S., S.N., R.A.H., E.R. and N.B.L. wrote the analysis pipeline.

**Author Information** The original data are available from the Spitzer Space Telescope archive, programs 30129 and 40685. Reprints and permissions information is available at www.nature.com/reprints. The authors declare no competing financial interests. Correspondence and requests for materials should be addressed to K.B.S. (kevin218@knights.ucf.edu).

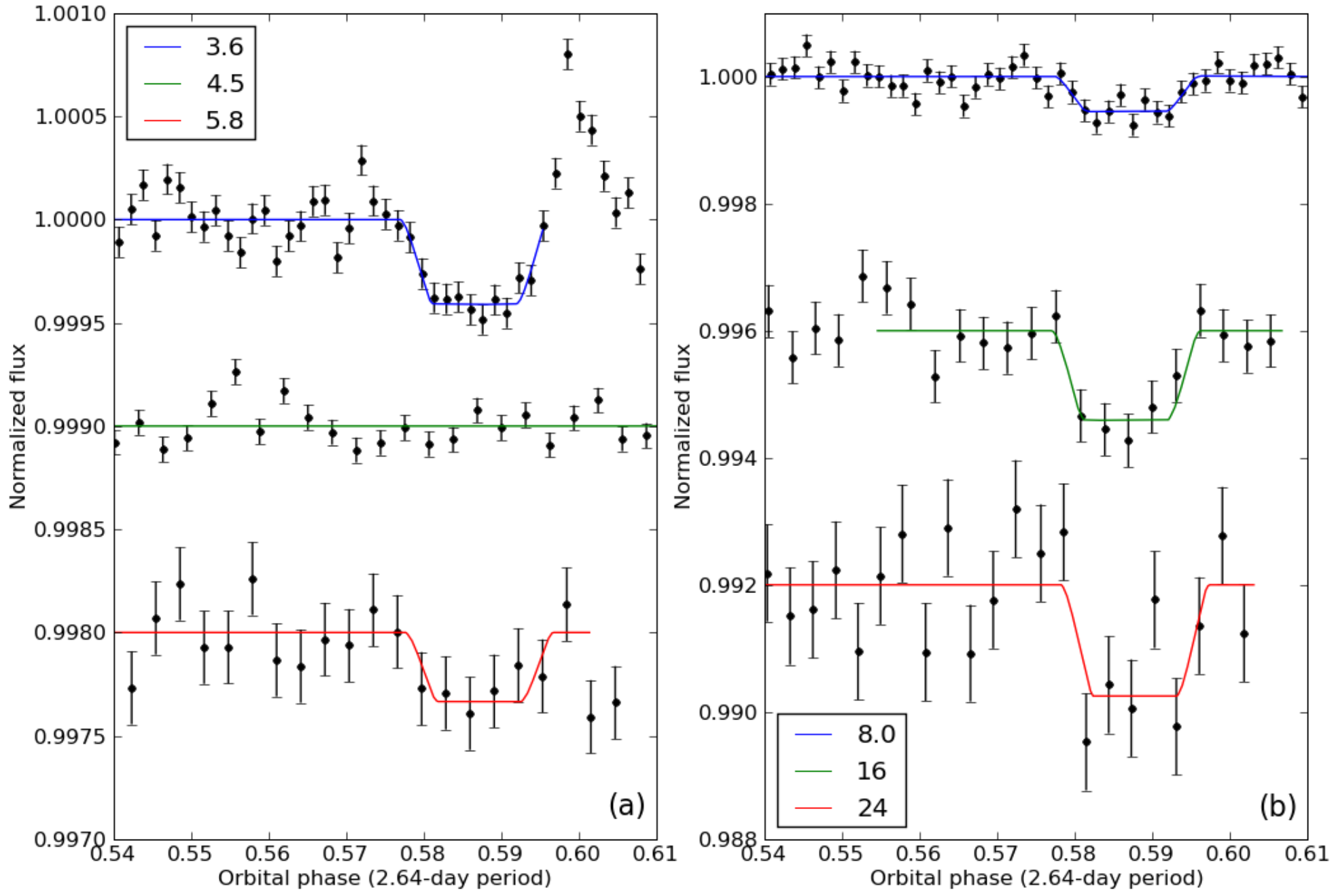

**Figure 1. Secondary eclipses of GJ 436b at six Spitzer wavelengths.** The flux values are corrected for sensitivity effects, normalized to the system brightness, and vertically separated for ease of comparison. **a**, Binned 3.6-, 4.5- and 5.8-μm data (with $1\sigma$ error bars), **b**, binned 8.0-, 16- and 24-μm data (with $1\sigma$ error bars), both with best-fit models and for orbital phases greater than 0.54. Note the different vertical scales used in each panel. The phase calculation uses an ephemeris time of Julian date 2,454,222.61588 and a period of 2.6438986 days (ref. 16). Because the planet passes behind the star, we ignore stellar limb darkening and use the uniform-source equations[29] for the eclipse shape. The position sensitivity model used either a quadratic[13] or a cubic function in the two spatial variables, including the cross terms to account for any correlation. An asymptotically constant exponential function[14] models the time-varying sensitivity. The 3.6- and 4.5-μm channels exhibit strong position sensitivity while the 5.8-μm channel reveals a weak correlation with pixel position. The unmodelled region at 3.6 μm may be the result of stellar activity[30]; a similar region at 5.8 μm is unmodelled for reasons presented in Supplementary

Information. We detect no eclipse at 4.5 µm, but constrain the flux modulation at its 3$\sigma$ upper bound by fixing the secondary eclipse phase to the mean weighted value of the other channels. We use asymptotically constant exponential and linear functions to model the time-varying sensitivities at 8.0 and 16 µm, respectively. Our 8.0-µm analysis agrees with prior analyses[3,4] but we obtain a slightly higher brightness temperature (Table 1) due, in part, to a more recent Kurucz model[24]. No correction is necessary at 24 µm.

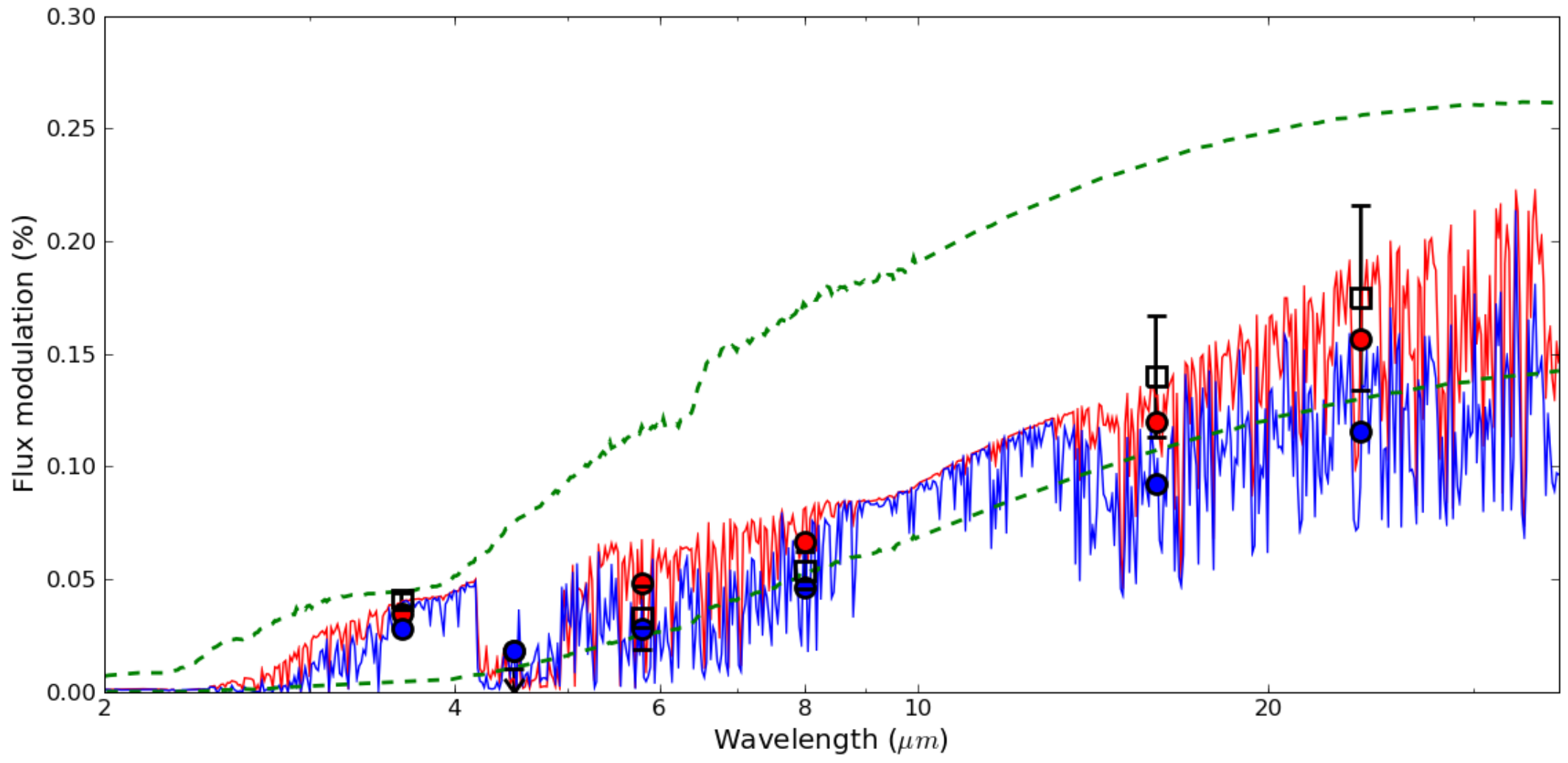

**Figure 2. Broadband spectrum constraints for GJ 436b**. The two atmospheric models (red and blue lines) have the same temperature structure and no thermal inversion. The red model has uniform mixing ratios for $H_2O$, $CH_4$, CO and $CO_2$ of $3 \times 10^{-6}$, $1 \times 10^{-7}$, $7 \times 10^{-4}$ and $1 \times 10^{-7}$, respectively. For the blue model, these are $1 \times 10^{-4}$, $1 \times 10^{-7}$, $1 \times 10^{-4}$ and $1 \times 10^{-6}$. The 3.6-µm channel is the key measurement in terms of constraining the methane abundance. It also limits the amount of $H_2O$ to less than that of CO, with little to no energy redistribution. Chemical equilibrium also requires some $NH_3$. The coloured circles are the bandpass-integrated models, the black squares are our data (with 1$\sigma$ error bars), and the black arrow depicts the 3$\sigma$ upper limit at 4.5 µm. The dashed green curves show blackbody spectra at 650 K (bottom) and 1,050 K (top) divided by the Kurucz stellar spectrum model[24]. The red and blue models have effective (equivalent blackbody) temperatures of 860 K and 790 K, respectively. We need not invoke an internal heat source[3]. Assuming zero albedo and planet-wide redistribution of heat, GJ 436b has an equilibrium temperature ($T_{eq}$, where emitted and absorbed radiation balance for an equivalent blackbody) of 770 K at periapse. For instantaneous reradiation of absorbed energy at secondary eclipse, the hemispheric effective temperature is 800 K and the peak temperature is 1,030 K.

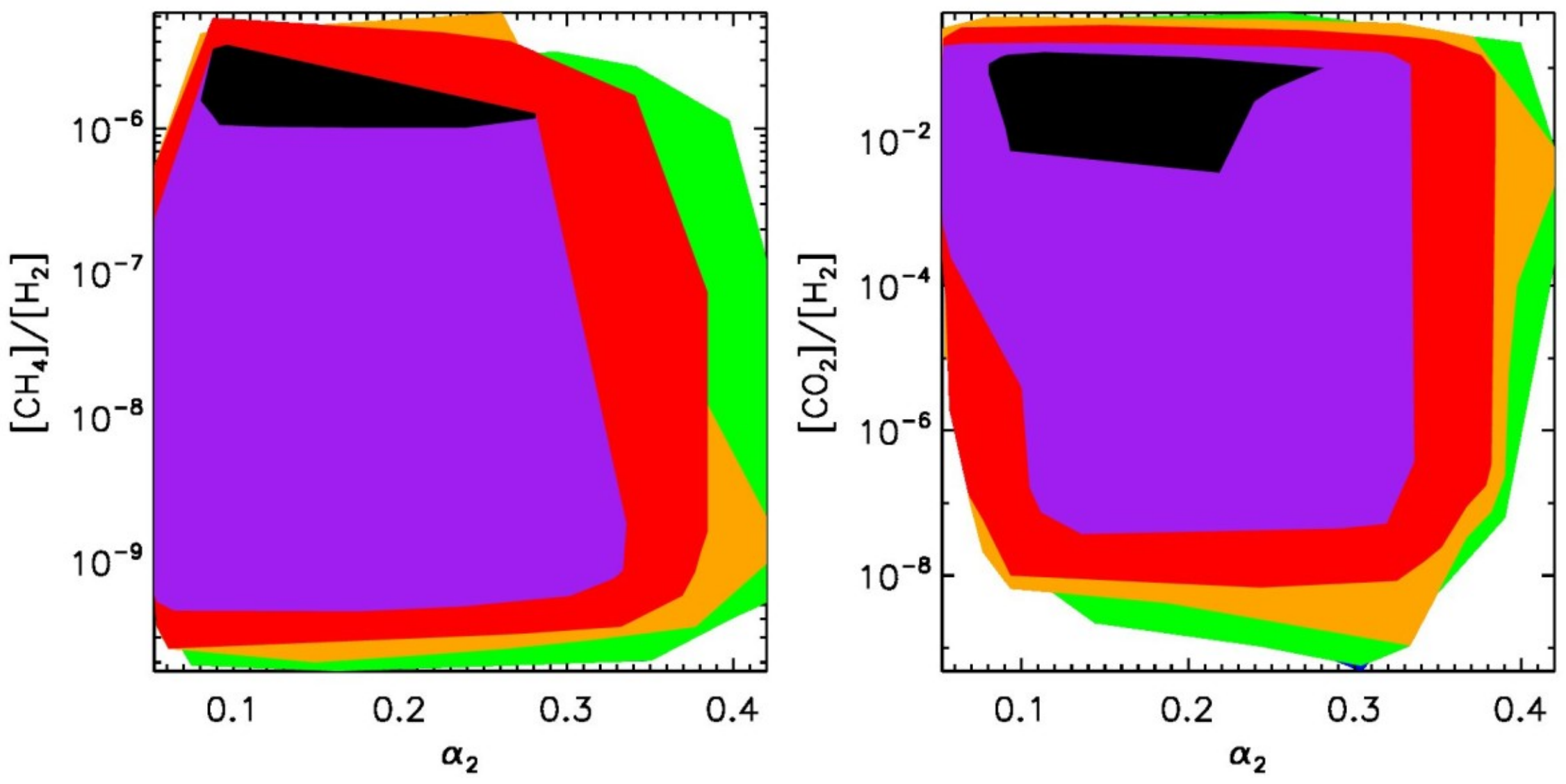

**Figure 3. Contours showing the explored mixing ratios of methane.** The purple, red, orange and green contours show error surfaces within $\xi^2$ of 1, 2, 3 and 4, where $\xi^2$ is $\chi^2$ divided by the number of channels. We use the $3\sigma$ upper limit for the 4.5-μm observation. The black surfaces show models with $\xi^2 < 1$ and $CH_4/H_2 > 10^{-6}$. The figure demonstrates that models with $CH_4$ mixing ratios close to $10^{-6}$ or above (panel **a**) require extremely large ($>10^{-3}$) $CO_2$ mixing ratios (panel **b**), which are unphysical based on current understanding of $CO_2$ photochemistry. The parameter space was explored with $\sim 10^6$ models. The $\alpha_2$ parameter, which is related to the temperature gradient in the lower atmosphere[18], was chosen arbitrarily for the abscissa.

**Table 1. Eclipse parameters and brightness temperatures**

| Channel (μm) | Eclipse midpoint (BJD − 2454000) | Eclipse duration (orbits) | Flux modulation (%) | Brightness temperature (K) |
|---|---|---|---|---|
| 3.6 | 496.4888 ± 0.0010 | 0.0192 ± 0.0008 | 0.041 ± 0.003 | 1,120 ± 20 |
| 4.5 | 499.1330 | 0.0191 | <0.010* | <700 |
| 5.8 | 501.778 ± 0.005 | 0.0191 | 0.033 ± 0.014 | 720 ± 110 |
| 8.0 | 282.3331 ± 0.0016 | 0.0186 ± 0.0014 | 0.054 ± 0.008 | 740 ± 40 |
| 16 | 477.981 ± 0.003 | 0.0191 ± 0.0023 | 0.140 ± 0.027 | 980 ± 130 |
| 24 | 470.053 ± 0.002 | 0.0191 | 0.175 ± 0.041 | 960 ± 170 |

BJD, barycentric Julian date; 1 orbit = 2.6438986 days. Eclipse duration is measured from start of ingress ($t_1$) to end of egress ($t_4$). Flux modulation is one minus in-eclipse flux divided by out-of-eclipse flux. Brightness temperature is the temperature of a similar blackbody that produces the same flux as the source in a given wavelength bandpass. The eclipses at 3.6, 8.0 and 16 μm are clear enough to fit durations; their weighted mean (72.6 ± 2.6 min) fixes the durations for the other wavelengths. We fix the ingress/egress times to 16 min (ref. 23) for all channels. Varying them produces equivalent times within the errors but degrades the overall fit quality. The Supplementary Tables and Supplementary Figures contain complete parameter results. The brightness temperature calculation[14] refers the flux modulation to a stellar spectrum model, which is interpolated from a grid of Kurucz models[24] using GJ 436's temperature (3,684 ± 71 K), log surface gravity (4.80 ± 0.10), and metallicity (−0.32 ± 0.12 dex)[2,25]. A Monte Carlo method computes the uncertainty in brightness temperature by varying the flux modulation and stellar parameters. The differing brightness temperatures at 3.6 and 4.5 μm suggest that these two wavelengths measure two different pressure levels; indeed Supplementary Fig. 5 shows that the 4.5-μm channel has an additional contribution from higher up in the atmosphere. To explain the 400 K difference in brightness temperatures, the model requires a very low concentration of methane.

*3$\sigma$ upper limit.

**Table 2. Atmospheric data for various planets**

| Planet | $H_2O$ | $CH_4$ | CO | $CO_2$ | $T_{eff}$ (K) | $CH_4$/CO |
|---|---|---|---|---|---|---|
| HD 209458b[18] (Spitzer broadband) | $\geq 10^{-8}$ $\leq 10^{-5}$ | $\geq 4 \times 10^{-8}$ $\leq 3 \times 10^{-2}$ | $\geq 4 \times 10^{-4}$ | $\geq 4 \times 10^{-9}$ $\leq 7 \times 10^{-8}$ | ≥1,310 ≤1,690 | $\gtrsim 10^{-7}$ $\lesssim 10^{2}$ |
| HD 189733b[18] (Spitzer broadband) | $\geq 10^{-5}$ $\leq 10^{-3}$ | $\leq 2 \times 10^{-6}$ | $\geq 7 \times 10^{-8}$ $\leq 2 \times 10^{-2}$ | $\geq 7 \times 10^{-7}$ $\leq 7 \times 10^{-5}$ | ≥1,480 ≤1,560 | $\gtrsim 10^{-10}$ $\lesssim 10$ |
| HD 189733b[26] (HST/NICMOS) | $\geq 1 \times 10^{-4}$ $\leq 1 \times 10^{-3}$ | $\leq 1 \times 10^{-7}$ | $\geq 1 \times 10^{-4}$ $\leq 3 \times 10^{-4}$ | $\geq 1 \times 10^{-6}$ $\leq 1 \times 10^{-5}$ | NA | $\lesssim 10^{-3}$ |
| GJ 436b (red) | $3 \times 10^{-6}$ | $1 \times 10^{-7}$ | $7 \times 10^{-4}$ | $1 \times 10^{-7}$ | 860 | $\sim 10^{-4}$ |
| GJ 570D[2] | $7 \times 10^{-4}$ | $5 \times 10^{-4}$ | $2 \times 10^{-6}$ | NA | 800 | $\sim 10^{2}$ |
| GJ 436b (blue) | $1 \times 10^{-4}$ | $1 \times 10^{-7}$ | $1 \times 10^{-4}$ | $1 \times 10^{-6}$ | 790 | $\sim 10^{-3}$ |
| Jupiter[27] | $\geq 2 \times 10^{-9}$ $\leq 2 \times 10^{-8}$* | $2.1 \times 10^{-3}$ | $1.6 \times 10^{-9}$ | $\leq 3 \times 10^{-9}$ | 110 | $\sim 10^{6}$ |
| Saturn[27] | $\geq 2 \times 10^{-9}$ $\leq 2 \times 10^{-8}$* | $4.5 \times 10^{-3}$ | $1 \times 10^{-9}$ | $3 \times 10^{-10}$ | 100 | $\sim 10^{6}$ |
| Uranus[28] | Ice | $2.3 \times 10^{-2}$ | $\leq 1.2 \times 10^{-8}$ | NA | 60 | $\sim 10^{6}$ |
| Neptune[28] | Ice | $2.9 \times 10^{-2}$ | $\sim 1 \times 10^{-6}$ | NA | 60 | $\sim 10^{4}$ |

Values given under headings $H_2O$, $CH_4$, CO and $CO_2$ are mixing ratios relative to hydrogen. $T_{eff}$, effective temperature; HST/NICMOS, Hubble Space Telescope Near Infrared Camera and Multi-Object Spectrometer. The planets are ordered in descending effective temperature. Chemical equilibrium predicts a roughly increasing $CH_4$-to-CO ratio. GJ 436b does not follow this general trend, as seen in the rightmost column. Its $CH_4$-to-$H_2$ mixing ratio is $>10^3$ times less than that of a brown dwarf of similar temperature and its $CH_4$-to-CO ratio is $>10^5$ times less. Excess CO may be the result of relatively strong vertical mixing[9]. A significant fraction of the methane may have polymerized into hydrocarbons[10], resulting in a shortage in observed $CH_4$. For comparison, GJ 436b's required methane mixing ratio of $10^{-7}$ is about $10^5$ times less than that on Uranus and Neptune, $10^4$ times less than that on Jupiter and Saturn, and ~20 times less than that on Earth, where methane is oxidized, not polymerized. NA, not available.

*Above cloud.

# Supplementary Information

At a relative flux level of just ~0.1% compared to the host star, exoplanet secondary eclipses are well below Spitzer's 2% relative photometric accuracy requirement[31]. This and their low intrinsic signal-to-noise ratios (SNR, often below 10) require that we attend closely to analysis details. Because different analysis approaches may obtain significantly different results, we also present more than the usual level of detail about our fits, so that future investigators who choose to analyze these data can compare their work to ours. This Supplementary Information (SI) presents how we determined the centres of the photometric apertures, adjusted for varying array sensitivity with respect to aperture centre location ("position sensitivity") and time ("ramp"), and fit models to the data. The final section presents the results of our fits in sufficient detail for evaluation of alternative analyses. Many other methods appear in the SI to Ref. 14.

## Centring and Photometry

Spitzer's instrumental point-spread functions (PSFs) are stable in time and vary little with the normal pointing wander (<1") over a few-hour staring observation. Since zodiacal light and instrumental effects contribute significant noise, we use a small aperture plus an aperture correction at short wavelengths and optimal photometry[15,32] at longer wavelengths. In either case, mismatching the aperture or PSF model to the data produces additional error, so one must determine PSF centres accurately. Here we compare three methods. The first[33] computes the centre of light of pixels within a circular aperture and above the frame's median value by at least 0.1% of the median-subtracted peak value. The second fits a two-dimensional (2D) Gaussian with free position parameters. The third, called least asymmetry, optimizes the stellar radial profile by calculating:

$$\text{Asym}\ (x, y) = \sum_{i=1}^{i_{\max}} n_{r_i} \sigma_{r_i}^4 , \qquad (1)$$

where $(x, y)$ is the current pixel location and $\sigma_r$ is the standard deviation of the $n_r$ pixels at the radial distance $r_i$ in pixels from the current central pixel. The first few discrete values for $r_i$ are 1, $\sqrt{2}$, 2, $\sqrt{5}$, $2\sqrt{2}$, 3, etc. We find using $i_{\max} = 5$ provides comparable precision and computes faster than larger values. An inverted 2D Gaussian with free position parameters finds the minimum in asymmetry space, which defines the centre of the object.

Tests using real datasets show that the 2D Gaussian and least-asymmetry methods are more precise than the centre-of-light method (see Supplementary Figure 1). For the example data, the Gaussian method is the most precise, but this is not always true. We tested the accuracy with a fake dataset made from a 100× oversampled Spitzer Point Response Function[34] (PRF) centred at 50 locations along a pixel diagonal. We rebinned each image to the nominal resolution, copied it 100 times, and added Gaussian noise. Supplementary Figure 2 plots the median residuals between the known and computed centres for the Gaussian and least-asymmetry methods. Both methods are comparable near the corners, but the least-asymmetry method is more accurate near a pixel centre. The Gaussian method is more consistent over the entire pixel range. For the observations of GJ 436b at 5.8 and 8.0 μm, the mean radial distances from their respective pixel centres are ~0.2 pixels, so, as indicated in Supplementary Table 1, the best centring method is least asymmetry. The evaluation metric is the standard deviation of the normalized (with respect to the stellar flux) residuals between the measured and model flux values.

The IRS and MIPS channels typically achieve their best results using optimal photometry, but 5×-interpolated aperture photometry[14] is best for the IRAC channels. Supplementary Table 1 gives the best aperture sizes, found by varying the size and minimizing the standard deviation of the normalized residuals. Changing the aperture size by 0.25 pixels from the best value increases this standard deviation by <0.4% and typically by much less, so smaller pixel increments are unnecessary.

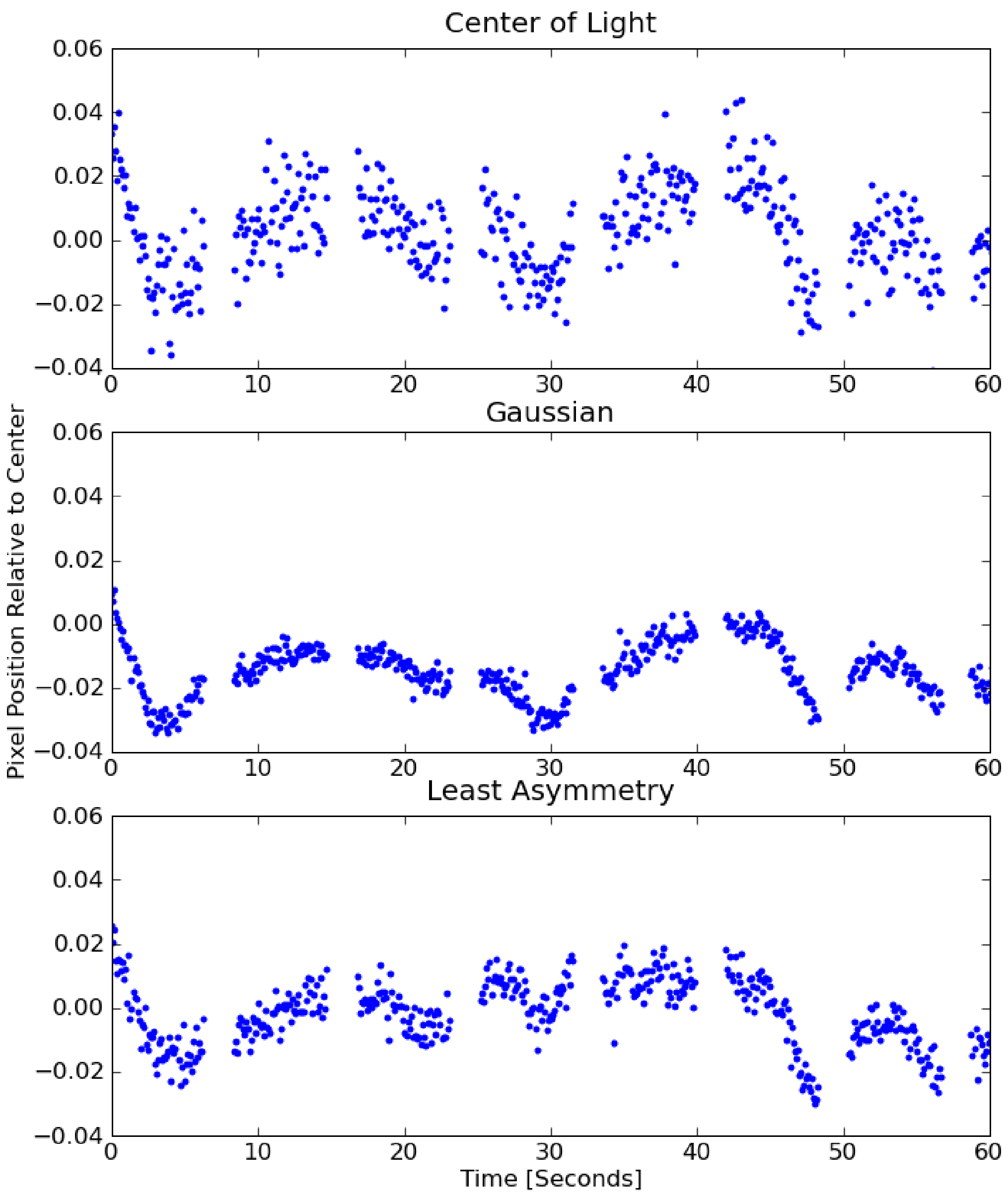

**Supplementary Figure 1. Three centring methods track the vertical position of GJ 436b for a small portion of the 3.6 µm data.** For this dataset, the Gaussian centring method most precisely tracks the spacecraft pointing. Small pointing oscillations occur on a ~5-second timescale. Gaps occur every 64 frames as the camera transfers data to the spacecraft's data system.

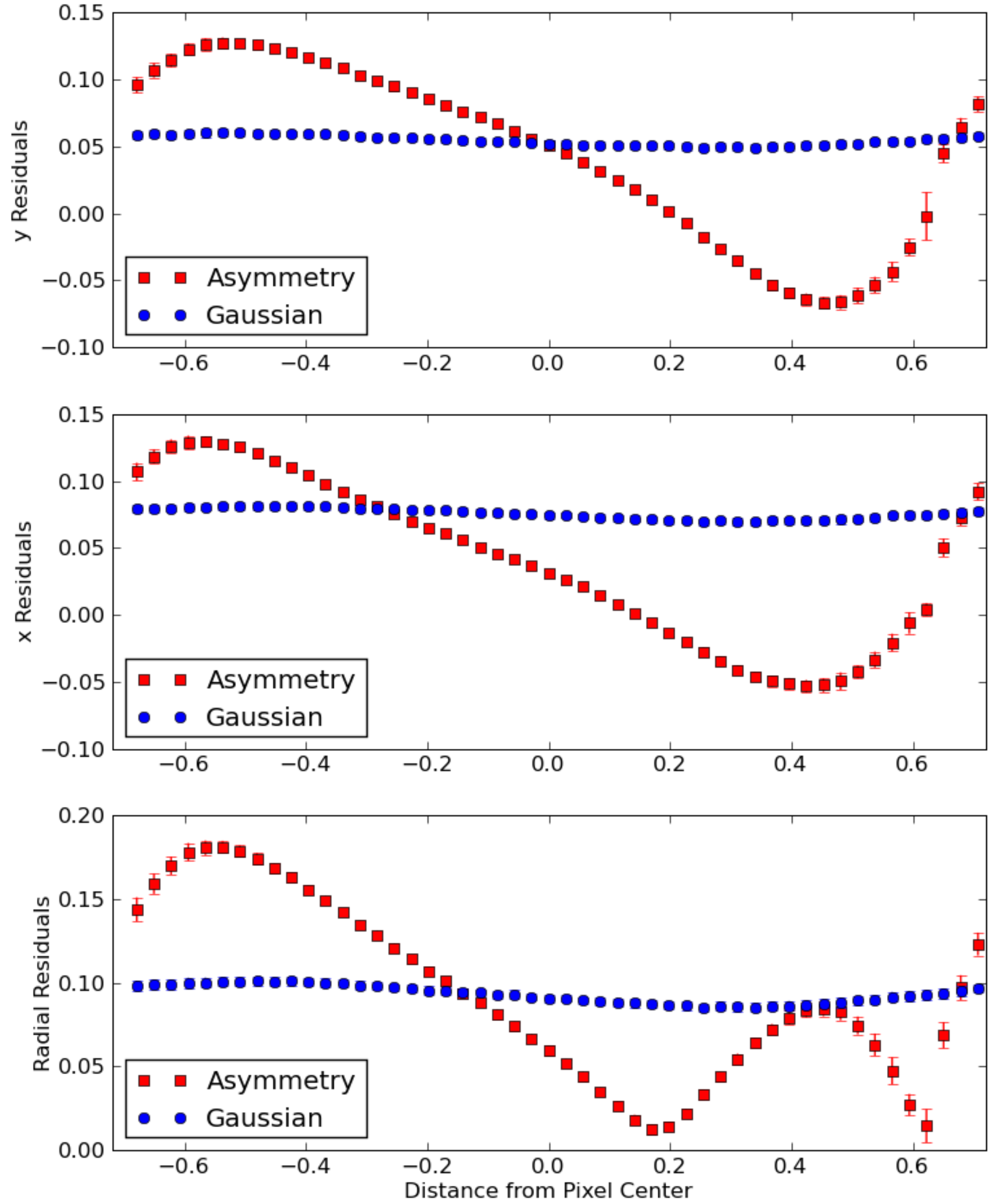

**Supplementary Figure 2. Comparison between Gaussian and least-asymmetry centring methods.** Each point ($1\sigma$ error bars) represents the median and standard deviation of the $y$ (top), $x$ (middle), and radial (bottom) residuals between the known and measured centroids using 100 synthetic PRF images, each with Gaussian noise, having true centres at the diagonal distances given along the abscissa. The least asymmetry method consistently outperforms the Gaussian method when the true centroid is close to the centre of the pixel. The Spitzer PRFs are not perfectly symmetric, resulting in the asymmetric form of the plots across the pixel centre.

**Supplementary Table 1. Centring method and photometry apertures.**

| Channel [μm] | Camera | Centring Method | Aperture Size [Pixels] |
|---|---|---|---|
| 3.6 | IRAC | Gaussian | 2.75 |
| 4.5 | IRAC | Gaussian | 4.75 |
| 5.8 | IRAC | Least Asymmetry | 3.25 |
| 8.0 | IRAC | Least Asymmetry | 3.75 |
| 16 | IRS | Gaussian | N/A |
| 24 | MIPS | Gaussian | N/A |

IRAC = Infrared Array Camera[31].
IRS = Infrared Spectrograph (blue peak-up array)[35].
MIPS = Multiband Imaging Photometer for Spitzer[36].

## Position Sensitivity

In the 3.6- and 4.5-μm Spitzer channels, sensitivity varies up to 3.5% with centroid position. We detect for the first time much smaller variations at 5.8 μm, which may be due to intra-pixel sensitivity variations or residual flat-field errors. Polynomial models in the two position variables fit the position sensitivity:

$$\varphi' = \varphi\left[ay^2 + bx^2 + cyx + dy + ex + 1\right] \tag{2}$$

$$\varphi' = \varphi\left[ay^3 + bx^3 + cy^2x + dyx^2 + ey^2 + fx^2 + gyx + hy + ix + 1\right] \tag{3}$$

where $\phi'$ and $\phi$ are the measured and corrected fluxes, respectively, $x$ and $y$ denote the PSF centre relative to the pixel centre, and $a – i$ are (potential) free parameters. In general (but not for these particular datasets), if many of the PSF centres fall on two or more pixels, the sensitivity difference between pixels (uncorrected flat field) becomes important. In this case, each of the visited pixels has its own correction.

## Time-Varying Sensitivity

Two functions model the time-varying sensitivity: an asymptotically constant exponential[14] and a combination of logarithmic plus linear functions (similar to Ref. 3):

$$\varphi' = \varphi \cdot [1 \pm \exp\{-a \cdot (t - t_0)\}] \qquad (4)$$

$$\varphi' = \varphi \cdot [a \cdot \ln(t - t_0) + b \cdot (t - t_0) + 1], \qquad (5)$$

where, in Eqn. 4, the positive and negative signs are used for exponentially decreasing and increasing variability, $t$ is the observation time, and the free parameters are $a$, $b$, and $t_0$. If both intra-pixel and time-varying sensitivities apply, their multiplied corrections use only one $\phi$. Although $\phi'$ in Eqn. 5 tends toward infinity at large $t$, this physical impossibility is not a problem for observations of a few hours. Eqn. 4 curves more, so it generally produces slightly deeper eclipses than Eqn. 5. Without any physical reason to prefer either function, we test both and report the one with the lowest Bayesian Information Criterion value (described below).

## Determining the Best Model

The Metropolis-Hastings algorithm, a specific Markov-Chain Monte Carlo (MCMC) method[12], explores the model phase space to estimate the values and uncertainties of the free parameters. The position sensitivity, time-varying sensitivity, and eclipse model elements evaluate simultaneously. The eclipse element has parameters for the phase of secondary eclipse (the fraction of one orbital period from mid-transit to mid-eclipse), the duration between the first and fourth contact points, the eclipse flux ratio (or modulation, one minus the in- versus out-of-eclipse flux values), the ingress and egress durations, and $\phi$ in the absence of any sensitivity model elements. These parameters define the shape of the eclipse following Ref. 29 for a uniform source. Spitzer Basic Calibrated Data (BCD) come with calculated flux uncertainties per pixel, which are typically too large[14]. After a "burn-in" of at least $10^5$ iterations to forget the starting conditions, we rescale the uncertainties to give a reduced $\chi^2$ of ~ 1. After $10^6$ or more iterations, the best-fit parameters are those with the least $\chi^2$ value. We calculate the 34$^{th}$ percentile in both directions from the median value to obtain uncertainties (averaged if close, quoted separately otherwise).

The Bayesian Information Criterion (BIC)[37,38] compares models with differing numbers of free parameters, heavily penalizing those with more, relative to the least $\chi^2$ method. The preferred model has the lowest BIC value:

$$\mathrm{BIC} = \frac{1}{\sigma^2}\sum_{i=1}^{n}\varepsilon_i^2 + k\cdot\ln\{n\}, \qquad (6)$$

where $\varepsilon_i$ and is the residual of the $i^{th}$ data point, $\sigma^2$ is the error variance, $n$ is the number of data points, and $k$ is the number of free parameters. Supplementary Table 2 lists the combinations of model elements used in each channel, the resulting standard deviation of the normalized residuals, and the BIC values. Position sensitivity terms that contribute negligibly to the fit are removed from the model. The type of position sensitivity model element used does not significantly affect the eclipse parameters but can reduce their uncertainties.

## Supplementary Discussion

The short-lived spike that occurred after the eclipse at 3.6 μm may be the result of stellar activity[30,39]. If this sharp increase in observed flux had affected the eclipse, the flux ratio would have been larger and the duration longer, thus requiring even lower levels of methane in the models and an inexplicably long duration. We contend that this is not the case and do not fit the affected points. The high interest in M-dwarf planets calls for observational study of M-dwarf activity, notably flares, across the spectrum.

**Supplementary Table 2. Eclipse free parameters and best models**

| Channel [μm] | Eclipse Free Parameters | Time-Varying Sensitivity | Position Sensitivity | Std. Dev. Of Norm. Residuals | BIC |
|---|---|---|---|---|---|
| 3.6 | Depth, Duration, Phase | - | Quadratic | 0.003839 | 100548 |
| | | | Cubic | 0.003830 | 100136 |
| 4.5 | Depth | Falling Exponential | Quadratic | 0.002449 | 37738 |
| | | | $x^2$, $x$ & $y$ terms only | 0.002450 | 37718 |
| 5.8 | Depth, Phase | Falling Exponential | - | 0.007208 | 35423 |
| | | | Quadratic | 0.007194 | 35335 |
| | | | $y^2$ term only | 0.007194 | 35293 |
| 8.0 | Depth, Duration, Phase | Rising Exponential | - | 0.004985 | 37802 |
| | | Log + Linear | - | 0.004984 | 37809 |
| 16 | Depth, Duration, Phase | Linear | - | 0.002939 | 875 |
| | | Quadratic | - | 0.002923 | 1022 |
| 24 | Depth, Phase | - | - | 0.006344 | 1179 |

The residuals are normalized to the stellar flux.

The last ~2500 photometry points (~5%) at 5.8 μm drop unexpectedly and are difficult to model. Including these values in the fit causes the best-fit flux ratio to decrease from 0.033% to 0.020% using the quadratic position sensitivity model. In addition, the eclipse phase changes drastically with the additional points, resulting in relatively large errors. The weaker flux ratio is comparable in magnitude to the remaining deviations from the model, attracting the Metropolis-Hastings algorithm to nearby local minima that mimic eclipses. Without the position sensitivity model, the best fit has a physically impossible negative flux ratio. By fixing the eclipse phase, as we did for the 4.5-μm photometry, the flux ratio histogram of the MCMC trials are Gaussian distributed (see below); however, the ramp curvature and phase offset parameters possess distinct bimodal distributions with standard deviations ~5 times larger than leaving the eclipse phase as a free parameter. We exclude these points from the final model.

The 5.8 and 8.0 µm channels use Si:As detectors and are not expected to have intra-pixel sensitivity variations like the In:Sb detectors for the 3.6- and 4.5-µm channels[13]. Nonetheless, the weak position sensitivity effect at 5.8 µm clearly improves the fit, as indicated by the lower BIC value in Supplementary Table 2 and as shown in Supplementary Figure 3. The oscillatory motion of the flux (top panel) is in phase with that of the position on the detector (bottom panel) and the best-fit curve mimics the flux motion with high precision. This may be due to intra-pixel sensitivity or uncorrected flat field errors. A possible micrometeoroid impact caused a sudden shift in position at phase = 0.58 (BJD = 2454501.7555). This did not affect the measured flux values, so we did not remove any frames from this event that were not already flagged as bad. There are small oscillations in the flux at 8.0 µm, but we find no correlation between flux and position.

The relative dependences of position sensitivity on the measured flux are apparent at the three lowest wavelengths in Supplementary Figure 4. The time-varying sensitivities[14] at 5.8, 8.0, and 16 µm are also evident. Previous analyses[3,4] at 8.0 µm used log plus linear and asymptotic exponential functions, respectively, to model the time-varying sensitivity. We use the latter, which typically results in slightly larger flux ratios compared to the log-plus-linear expression. The pixel sensitivity at 16 µm increases by ~1.5% until the phase reaches 0.54. It then stabilizes before decreasing in sensitivity. We only model the decreasing section, using a linear function. Other models produced larger BIC values. The mean images in the MIPS dataset, with bad pixels removed, revealed a clear, roughly quadratic rise in the background level along the $y$ axis. This effect varied with position but was consistent at each scan mirror tilt angle. We thus subtracted the median value along the $x$-axis from each row of each image. However, the photometric results from the background-subtracted images did not show improvement, so we used the uncorrected data.

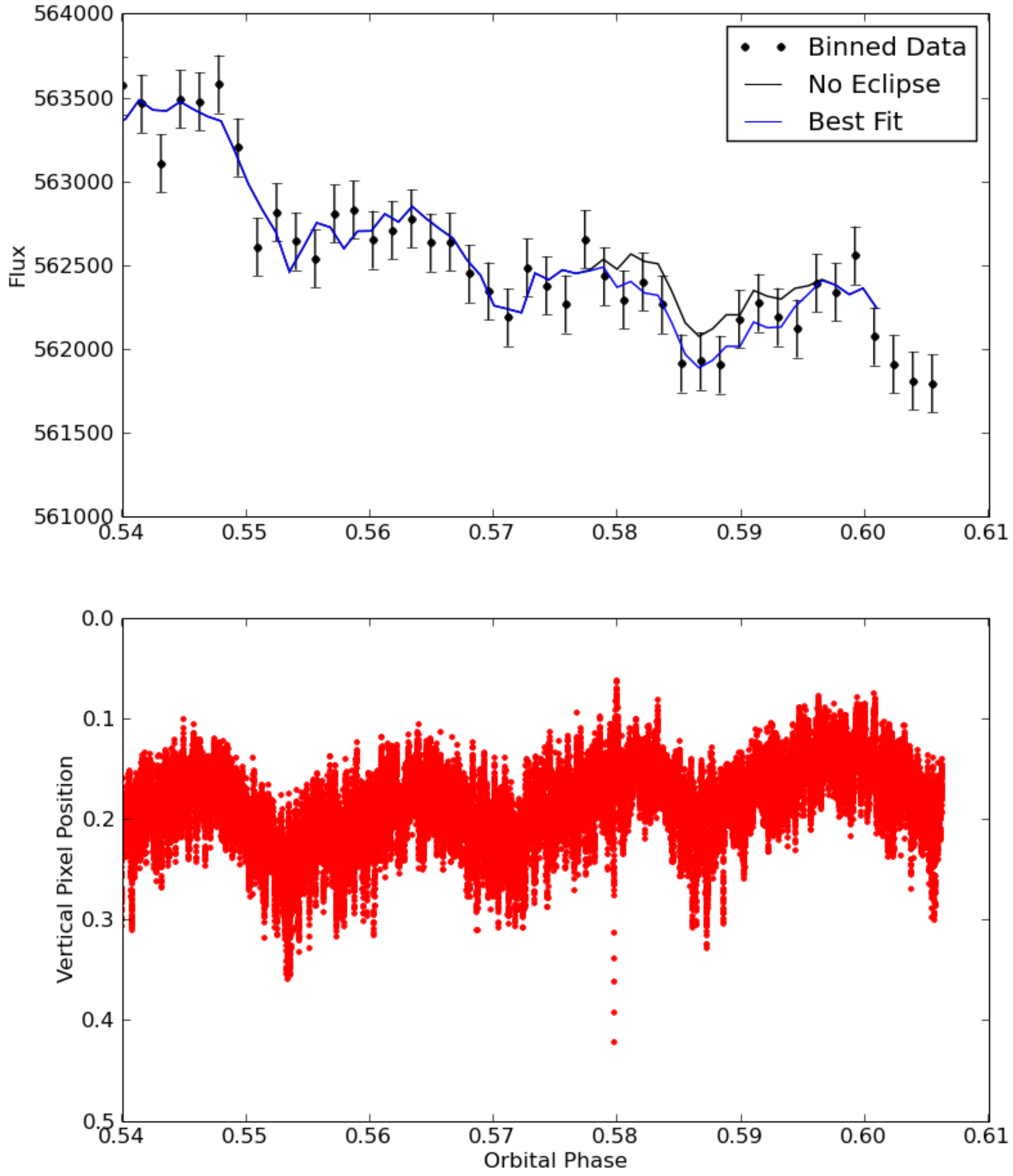

**Supplementary Figure 3**. **Position sensitivity at 5.8 μm.** The top panel plots the binned fluxes and best-fit model *vs.* phase. The bottom panel shows the unbinned vertical pixel positions (least asymmetry method, Gaussian is similar), which correlate with the measured flux values. Note the position excursion – possibly a micrometeoroid hit – at phase ~0.58.

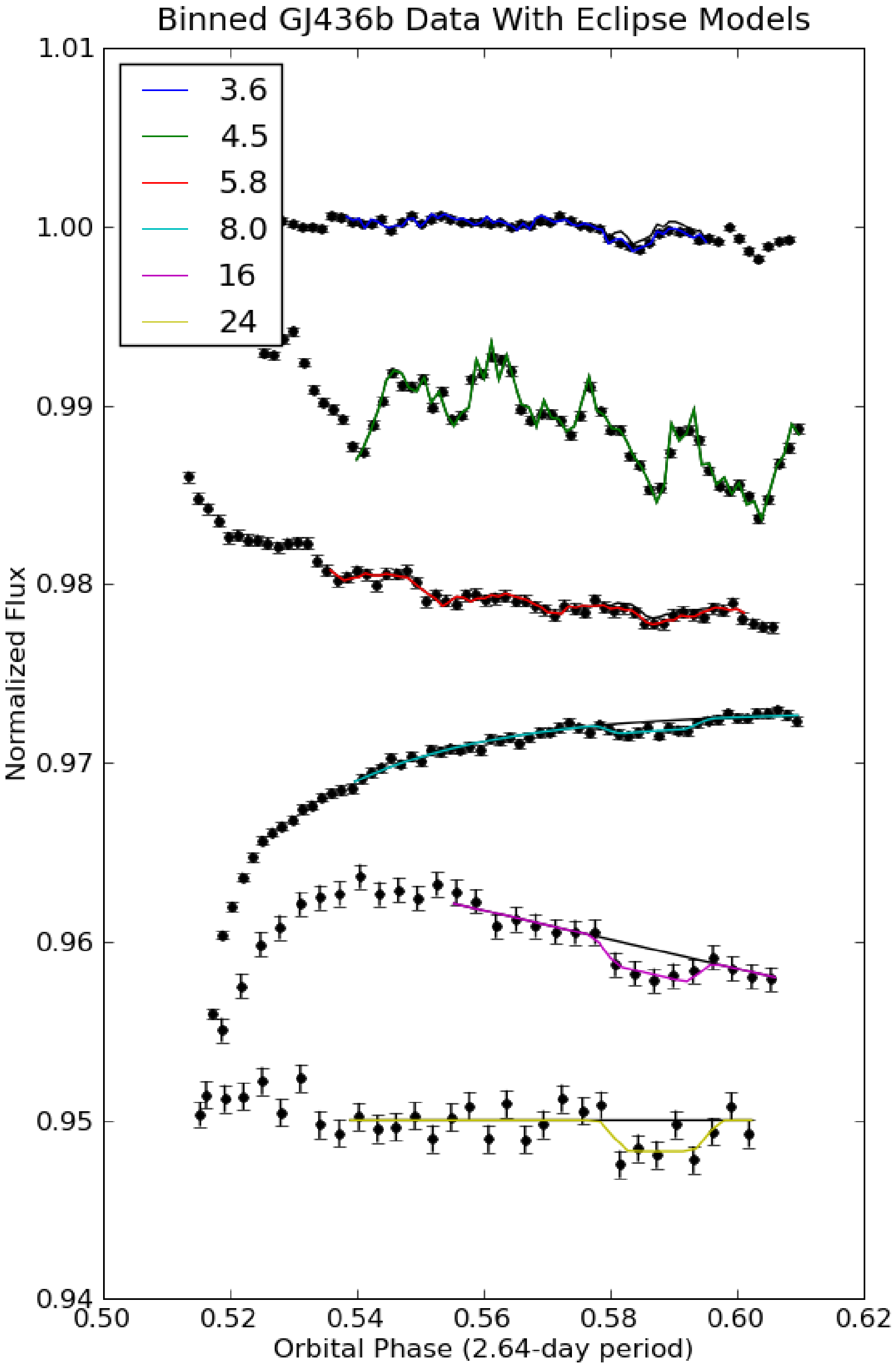

**Supplementary Figure 4. Binned, normalized, raw photometry of the GJ 436 system in all six channels with eclipse and systematic models.** The channels are vertically offset for clarity. The black curves do not include the eclipse model elements. At 4.5 μm, the eclipse depth is too small to distinguish.

# Supplementary Tables

**Supplementary Table 3. Best-fit orbital parameters with corresponding errors.**

| Parameter | Best Fit | Error |
|---|---:|---:|
| Period (Days) | 2.6438983 | ± 0.0000016 |
| Ephemeris Time (JD) | 2454222.61587 | ± 0.00012 |
| Argument of Periapsis (°) | 357 | ± 10. |
| Eccentricity | 0.1371 | + 0.0048<br>- 0.00013 |
| Semi-Amplitude (m/s) | 18.2 | ± 0.4 |
| Linear Slope (m/s/yr) | 1.27 | ± 0.20 |
| Linear Offset (m/s) | 4.1 | ± 0.7 |

We used published transit[16] and RV[17] data but removed two points due to the Rossiter-McLaughlin effect[40]. Our MCMC orbit routine fit the period, ephemeris time, argument of periapsis ($\omega$), eccentricity ($e$), semi-amplitude ($K$), a linear correction slope ($\mathrm{d}v/\mathrm{d}t$), and an offset ($\gamma$) term.

**Supplementary Table 4. Best-fit free parameters at 3.6 μm.**

| Parameter | Best Fit | Low Error | High Error | SNR |
|---|---|---|---|---|
| Eclipse Phase [orbits] | 0.5867 | -0.0004 | 0.0004 | 1,600 |
| Eclipse Duration [orbits] | 0.0192 | -0.0008 | 0.0008 | 23.0 |
| Flux Ratio [%] | 0.041 | -0.003 | 0.003 | 12.1 |
| Star Flux [μJy] | 1,287,800 | -500 | 600 | 2,350 |
| Intra-pixel, Cubic Term in y | 0.11 | -0.02 | 0.02 | 5.3 |
| Intra-pixel, Cubic Term in x | -0.057 | -0.004 | 0.004 | 12.8 |
| Intra-pixel, $y^2x$ Cross Term | 0.12 | -0.02 | 0.04 | 3.9 |
| Intra-pixel, $yx^2$ Cross Term | 0.185 | -0.035 | 0.014 | 7.5 |
| Intra-pixel, Quadratic Term in y | -0.710 | -0.04 | 0.05 | 15.1 |
| Intra-pixel, Quadratic Term in x | -0.0200 | -0.0020 | 0.0017 | 10.9 |
| Intra-pixel, yx Cross Term | -0.011 | -0.006 | 0.007 | 1.7 |
| Intra-pixel, Linear Term in y | -0.058 | -0.005 | 0.009 | 8.6 |
| Intra-pixel, Linear Term in x | 0.0127 | -0.0010 | 0.0010 | 12.4 |

**Supplementary Table 5. Best-fit free parameters at 4.5 μm.**

| Parameter | Best Fit | Low Error | High Error | SNR |
|---|---|---|---|---|
| Flux Ratio [%] | 0.0002 | -0.0032 | 0.0034 | 0.075 |
| Star Flux [μJy] | 861,900 | -200 | 300 | 3,470 |
| Ramp,Curvature | 29.04 | -0.08 | 0.11 | 307 |
| Ramp,Phase Offset | 0.281 | -0.004 | 0.003 | 76.7 |
| Intra-pixel, Quadratic Term in x | 0.083 | -0.003 | 0.004 | 22.7 |
| Intra-pixel, Linear Term in y | 0.1471 | -0.0006 | 0.0005 | 267 |
| Intra-pixel, Linear Term in x | 0.0747 | -0.0017 | 0.0022 | 37.7 |

### Supplementary Table 6. Best-fit free parameters at 5.8 μm.

| Parameter | Best Fit | Low Error | High Error | SNR |
|---|---|---|---|---|
| Eclipse Phase [orbits] | 0.5873 | -0.0042 | 0.0016 | 202 |
| Flux Ratio [%] | 0.033 | -0.015 | 0.014 | 2.3 |
| Star Flux [μJy] | 562,190 | -190 | 230 | 2,690 |
| Ramp, Curvature | 22.8 | -1.2 | 2.2 | 13.3 |
| Ramp, Phase Offset | 0.293 | -0.019 | 0.022 | 13.3 |
| Intra-pixel, Quadratic Term in y | -0.032 | -0.003 | 0.003 | 12.0 |

### Supplementary Table 7. Best-fit free parameters at 8.0 μm.

| Parameter | Best Fit | Low Error | High Error | SNR |
|---|---|---|---|---|
| Eclipse Phase [orbits] | 0.5867 | -0.0006 | 0.0006 | 955 |
| Eclipse Duration [orbits] | 0.0186 | -0.0014 | 0.0015 | 12.9 |
| Flux Ratio [%] | 0.054 | -0.008 | 0.008 | 7.3 |
| Star Flux [μJy] | 305,464 | -16 | 16 | 19,500 |
| Ramp, Curvature | 41.69 | -0.18 | 0.12 | 278 |
| Ramp, Phase Offset | 0.4068 | -0.0008 | 0.0008 | 505 |

### Supplementary Table 8. Best-fit free parameters at 16 μm.

| Parameter | Best Fit | Low Error | High Error | SNR |
|---|---|---|---|---|
| Eclipse Phase [orbits] | 0.5866 | -0.0011 | 0.0009 | 588 |
| Eclipse Duration [orbits] | 0.0191 | -0.0026 | 0.0020 | 8.2 |
| Flux Ratio [%] | 0.140 | -0.025 | 0.029 | 5.3 |
| Star Flux [μJy] | 85,949 | -14 | 15 | 5,880 |
| Ramp, Linear Term | -0.082 | -0.008 | 0.006 | 11.6 |

**Supplementary Table 9. Best-fit free parameters at 24 μm.**

| Parameter | Best Fit | Low Error | High Error | SNR |
|---|---|---|---|---|
| Eclipse Phase [orbits] | 0.5878 | -0.0008 | 0.0008 | 747 |
| Flux Ratio [%] | 0.175 | -0.042 | 0.041 | 4.2 |
| Star Flux [μJy] | 38,017 | -7 | 7 | 5,310 |

## Supplementary Figures

Supplementary Figure 5 presents the contribution functions[41] and temperature profile *vs*. pressure (or depth) for all six observed channels. Supplementary Figures 6 - 11 present histograms of the free parameter values in the MCMC chains. To remove the correlation of the steps, the plots include only a fraction of the values plotted. For the low S/N datasets such as 4.5 and 5.8 μm, the chains explore physically impossible negative eclipse depths in order to ascertain the error. Most of the histograms are roughly Gaussian in shape but some parameters exhibit non-Gaussian errors.

Supplementary Figures 12 - 17 show correlations between free parameters in a small (for clarity) but representative percentage of the Markov steps. The MCMC random walk does not always produce smooth distributions. Outlier clumps can occur where the phase space has nearby local minima. Narrow paths can result from an ergodic probability distribution, which can reach any point in the bounded phase space. The eclipse parameters are generally uncorrelated with the intra-pixel and time-varying sensitivity parameters. However, strong correlations do occur between the star flux and certain intra-pixel terms and amongst the intra-pixel terms themselves. Due to the form of Eqns. 2 and 3, we expect some degree of correlation.

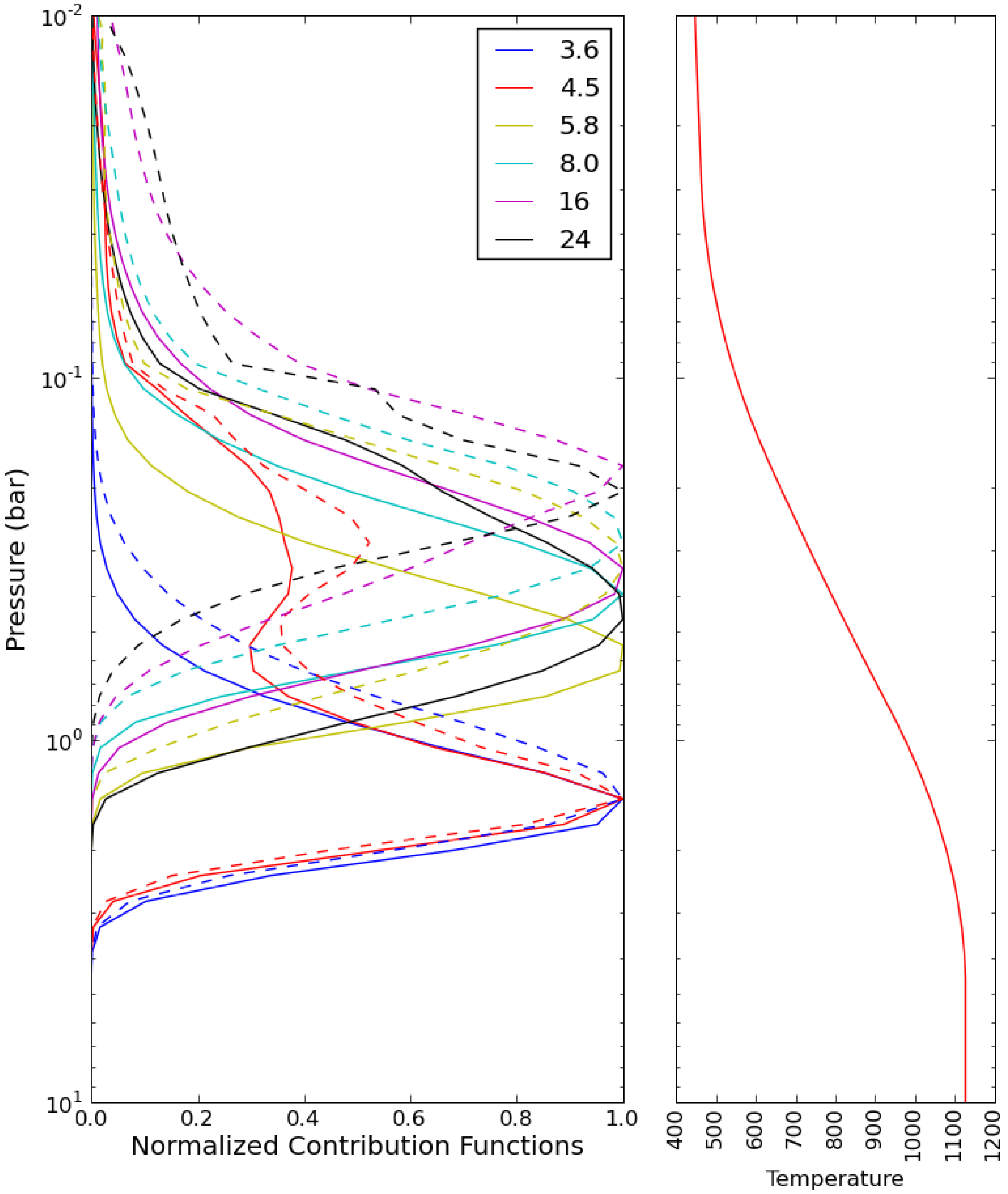

**Supplementary Figure 5. Normalized contribution functions[41] of GJ 436b in all six observed channels (left) and the corresponding temperature profile (right).** In the left frame, the solid lines are from the red model of Figure 2 (main paper); the dashed lines are from the blue model.

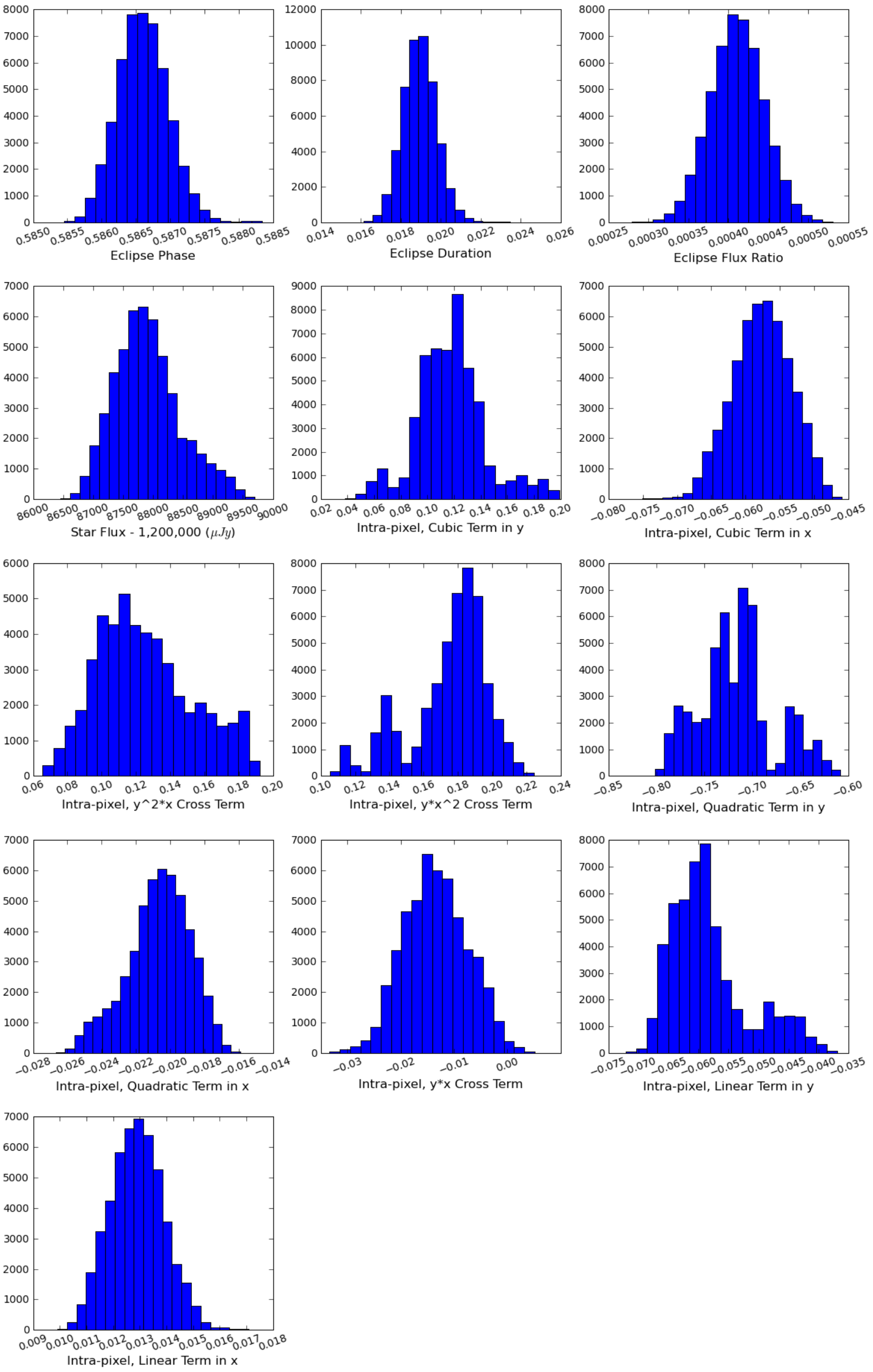

**Supplementary Figure 6. Histograms of free parameters at every 100th MCMC step (out of $5{\times}10^6$) at 3.6 μm.** The intra-pixel effect is most sensitive along the *y* axis, with the $y^2$ term dominating.

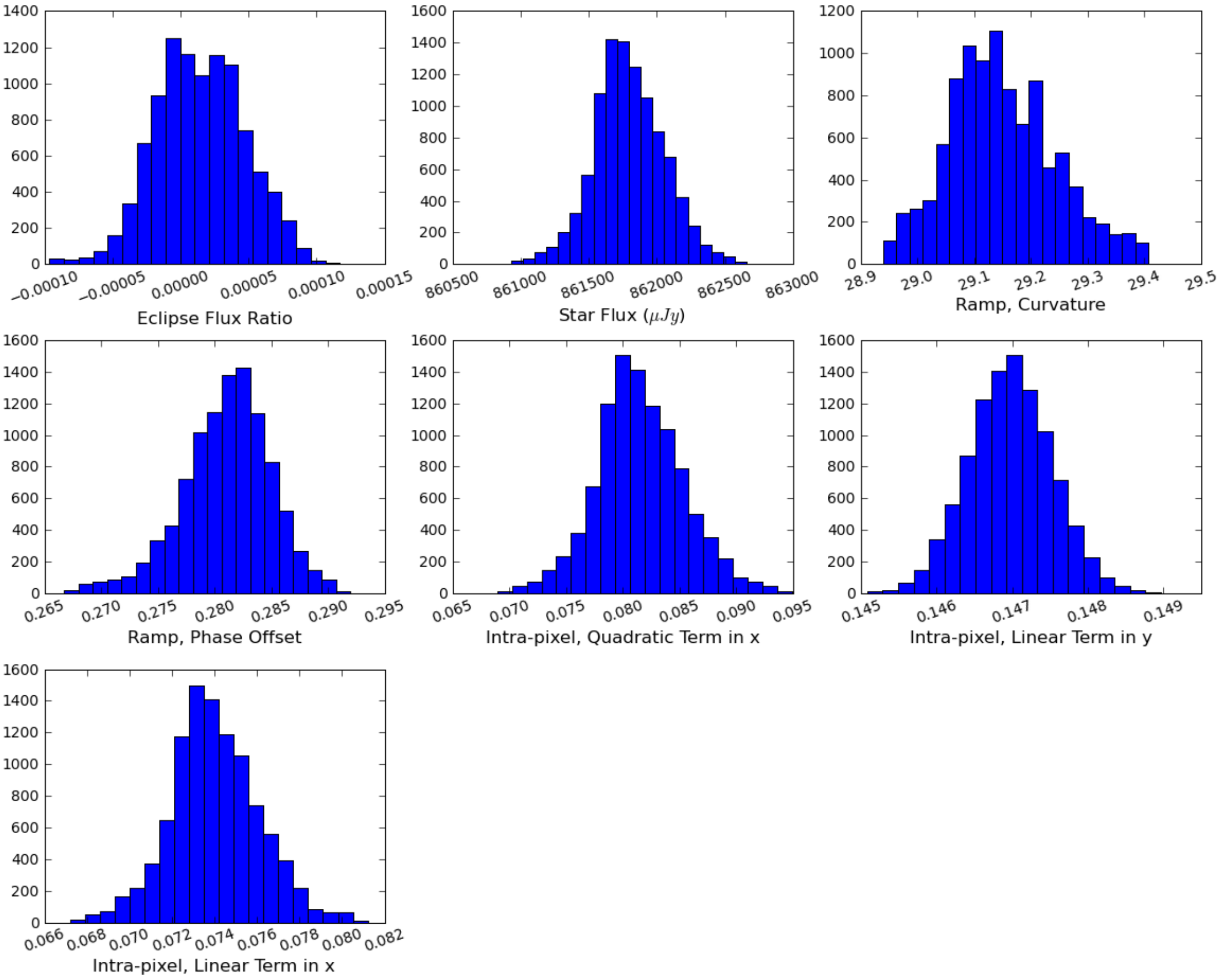

**Supplementary Figure 7. Histograms of free parameters at every 100$^{th}$ MCMC step (out of 10$^{6}$) at 4.5 μm.** The *y* term is the most dominant intra-pixel term at this wavelength.

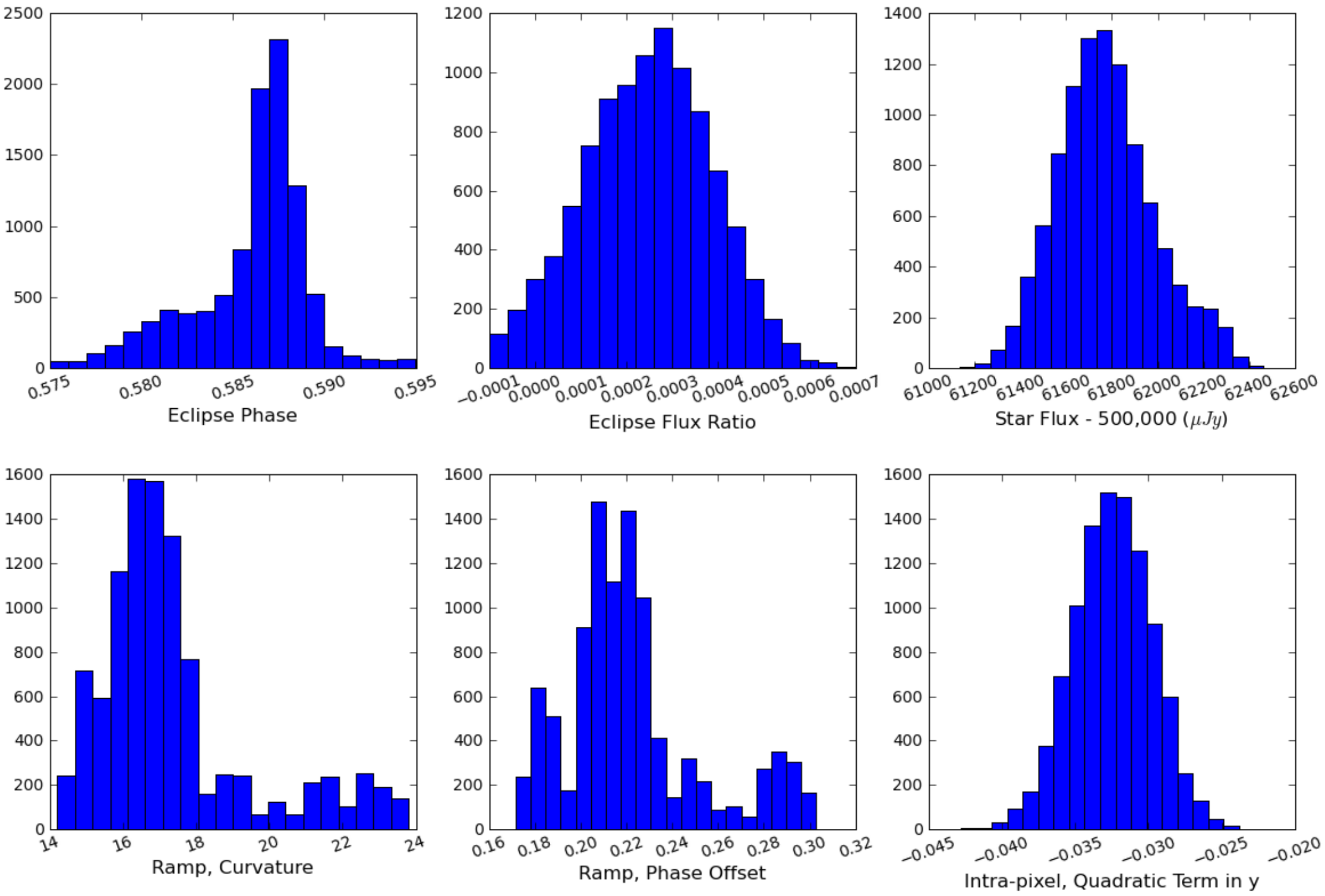

**Supplementary Figure 8. Histograms of free parameters at every 100$^{th}$ MCMC step (out of 10$^6$) at 5.8 µm.** Only the *y* axis intra-pixel dependence is significant.

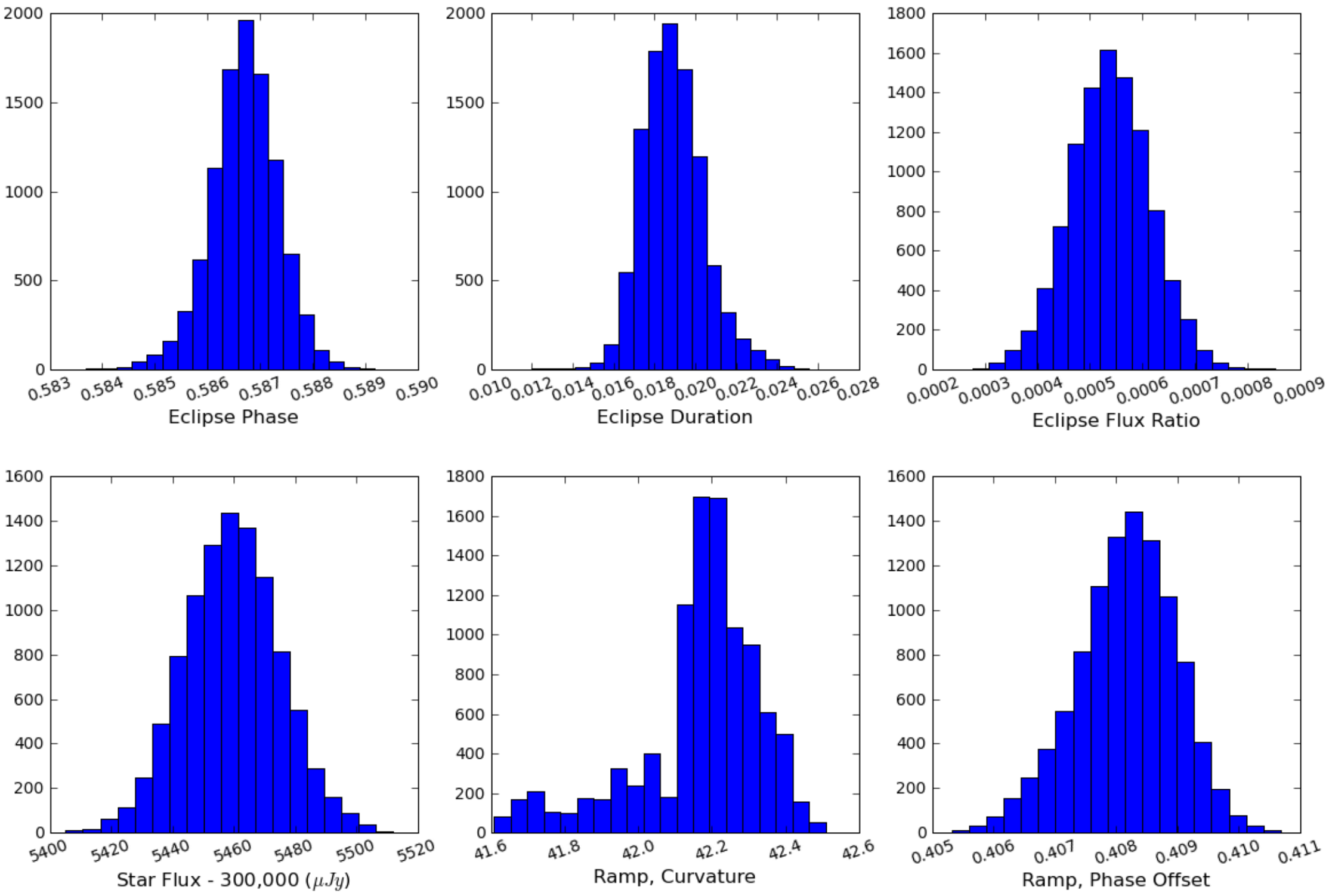

**Supplementary Figure 9. Histograms of free parameters at every 100th MCMC step (out of $10^6$) at 8.0 μm.**

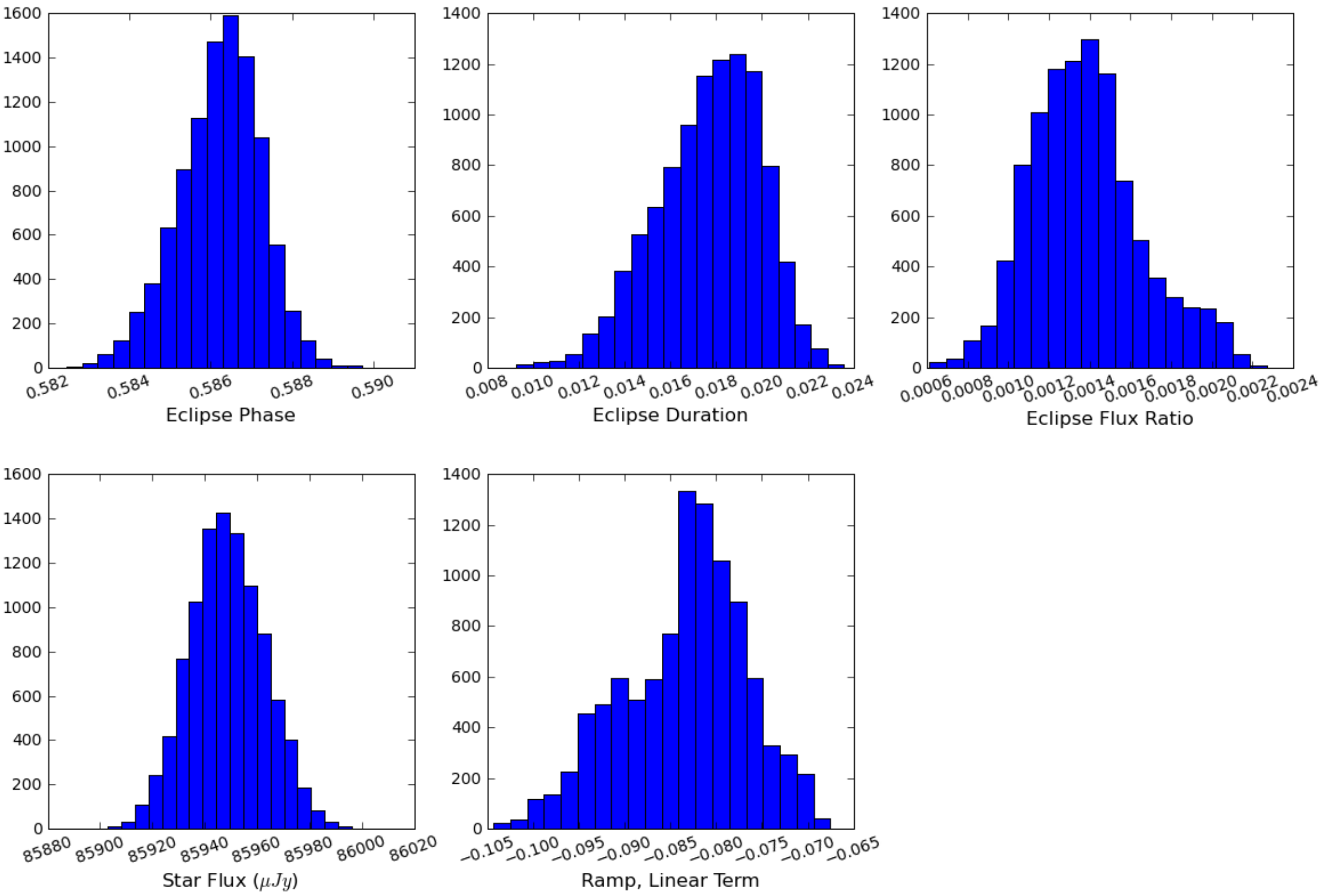

**Supplementary Figure 10. Histograms of free parameters at every 100$^{th}$ MCMC step (out of $10^6$) at 16 μm.**

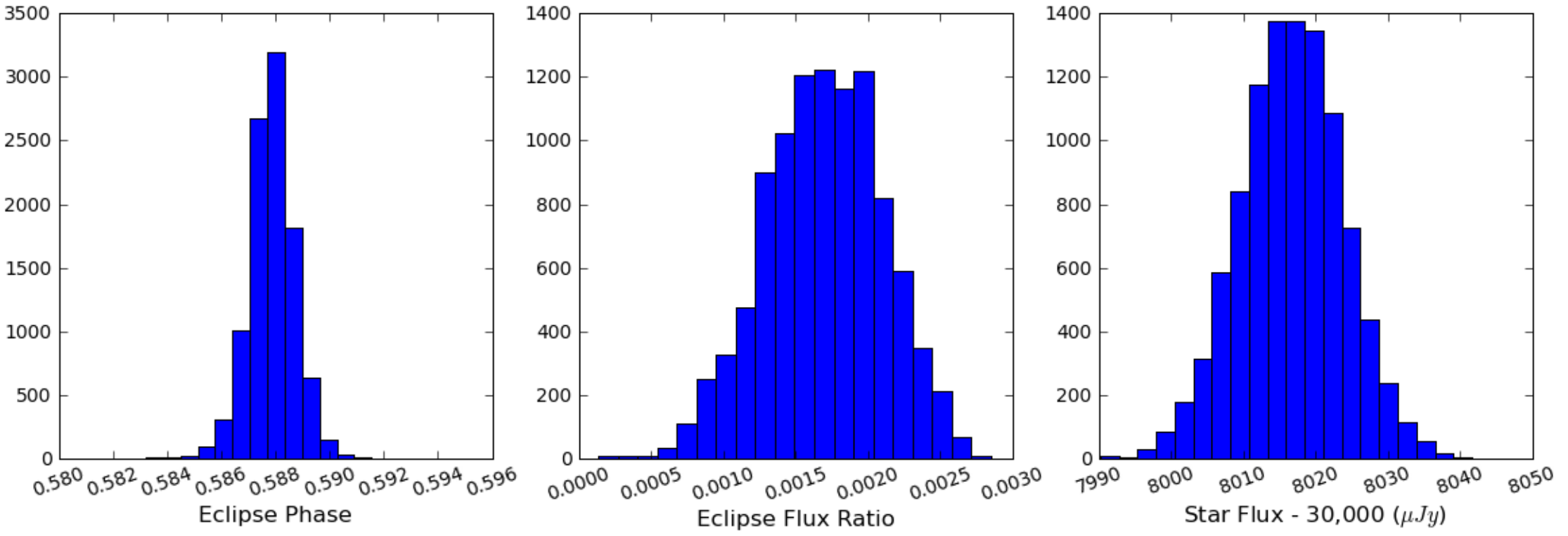

**Supplementary Figure 11. Histograms of free parameters at every 100$^{th}$ MCMC step (out of $10^6$) at 24 μm.**

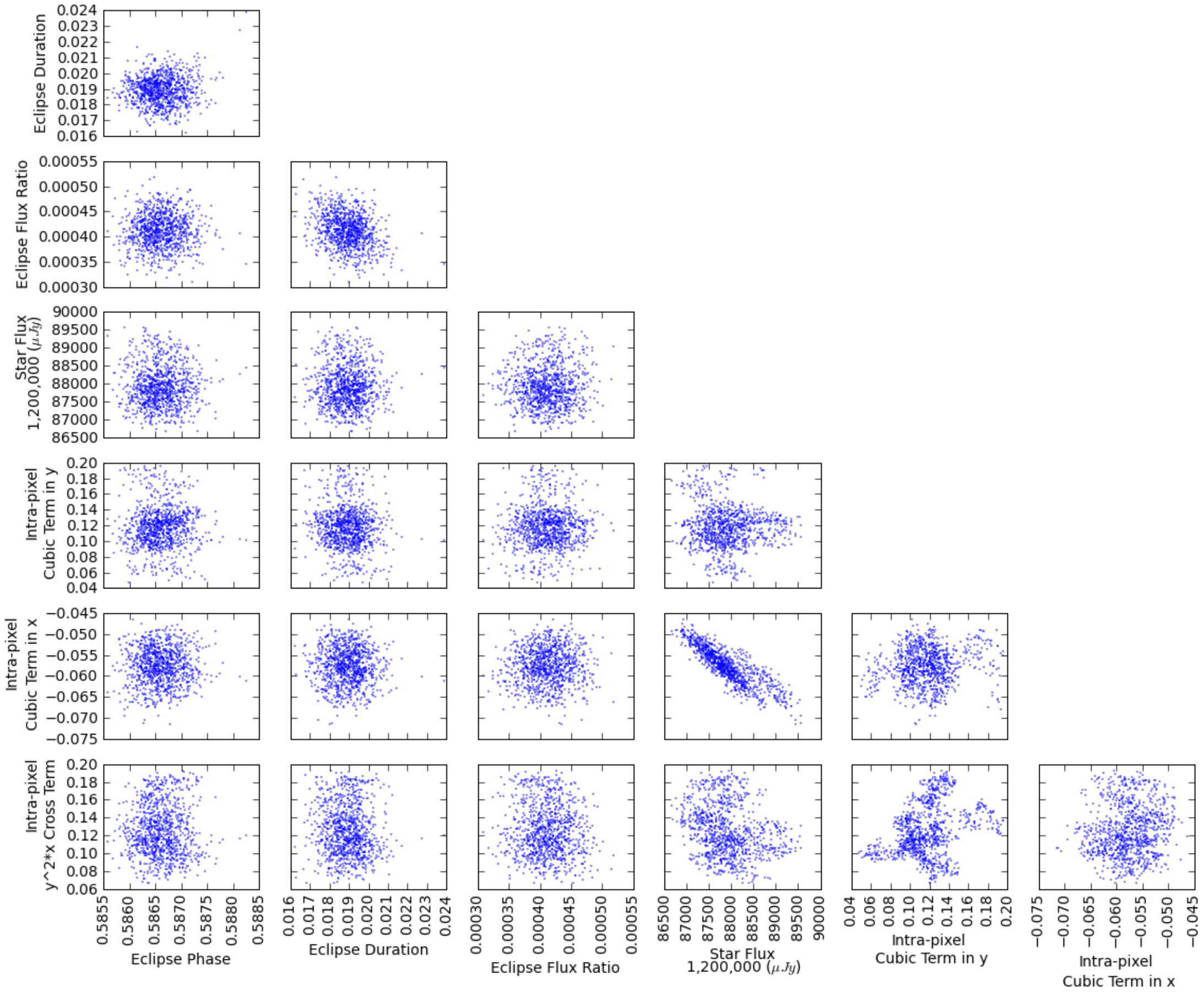

**Supplementary Figure 12a. See description below.**

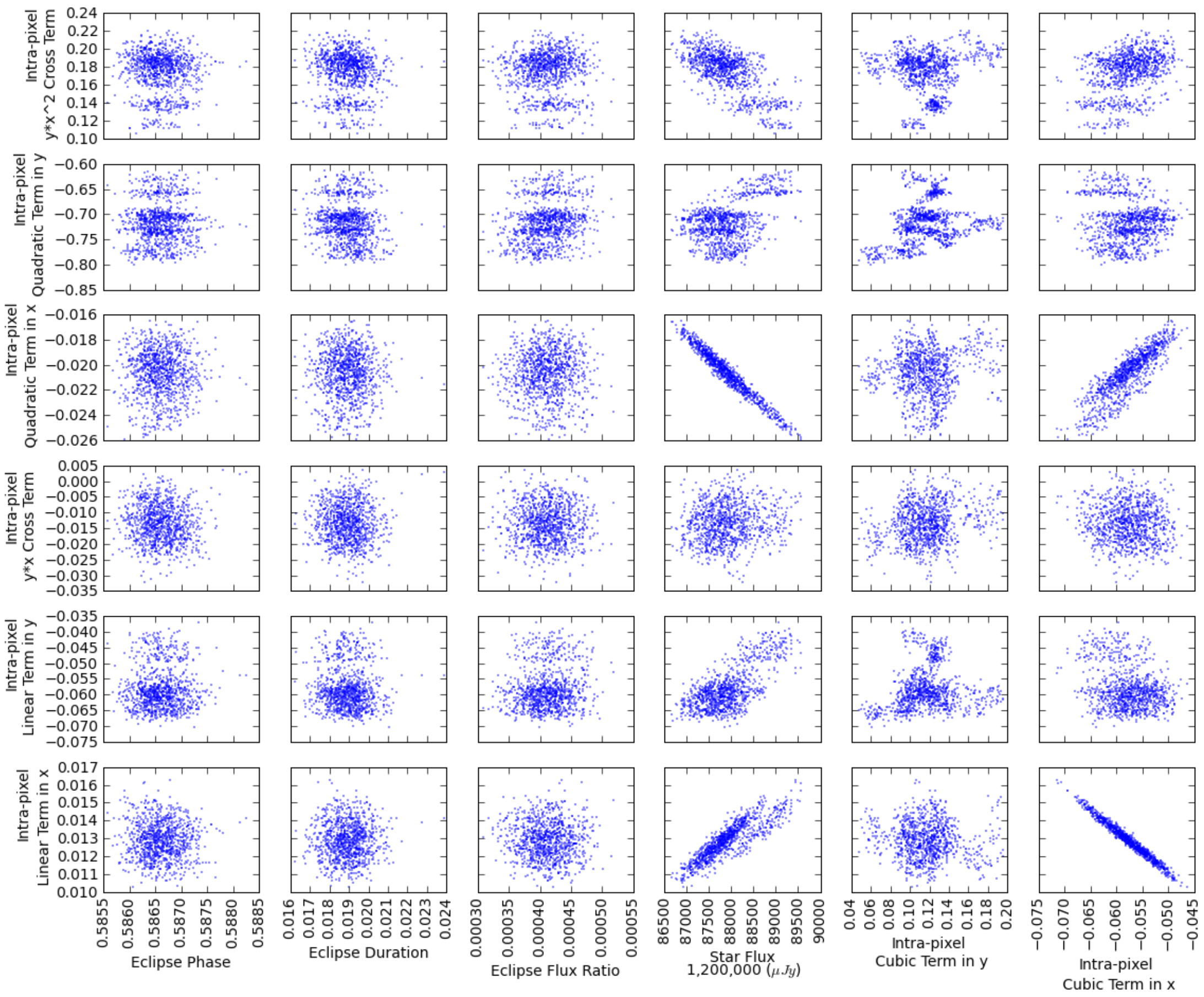

**Supplementary Figure 12b. See description below.**

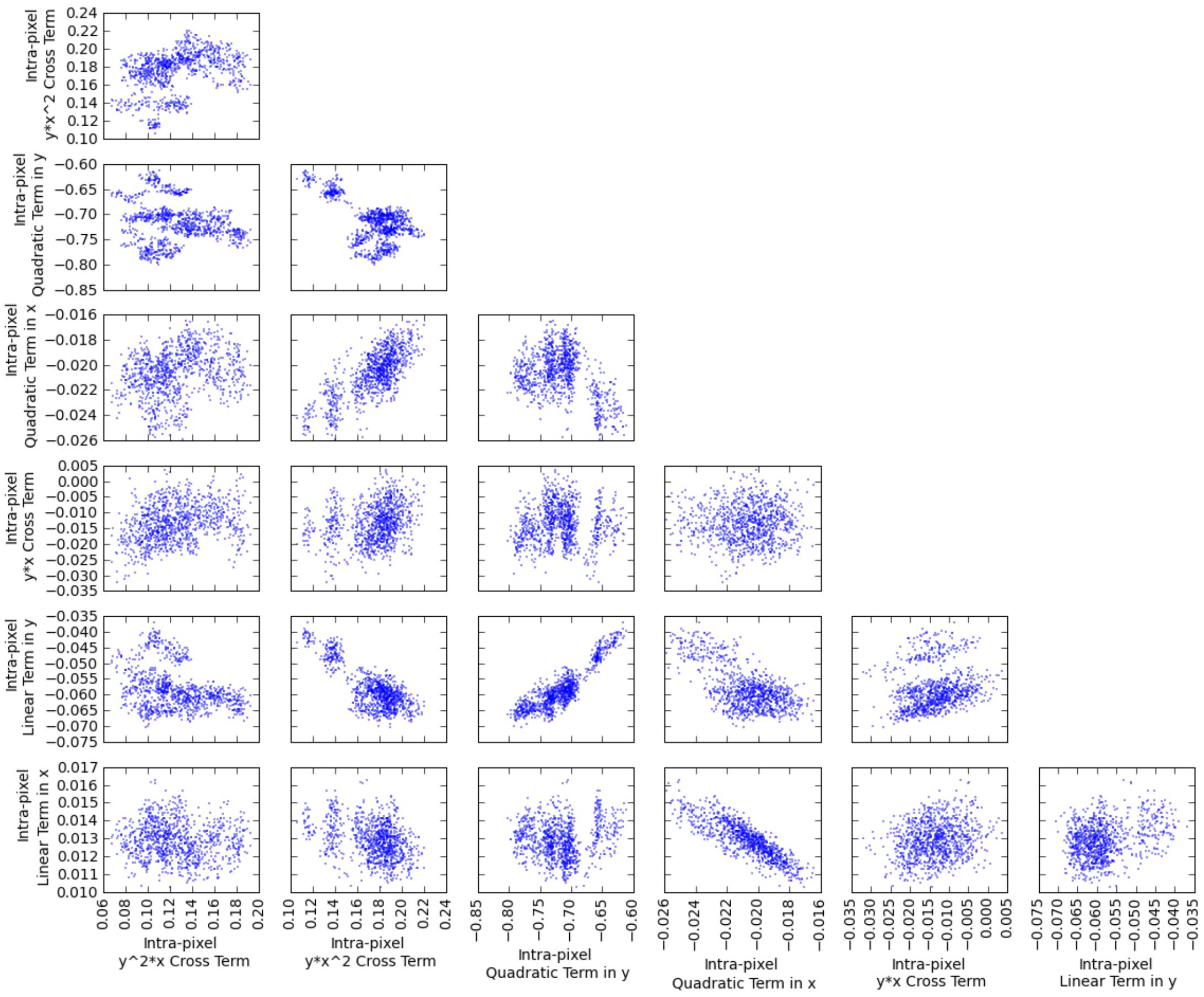

**Supplementary Figure 12c.**

**Supplementary Figure 12. Phase-space projections for every 1000$^{th}$ MCMC step at 3.6 µm.** Due to the large number of free parameters in this particular model, the phase-space projections are subdivided into three figures, labeled 12a, 12b, and 12c. The $y$ and $y^2$ terms are strongly correlated, with a coefficient of 0.94. The $x$, $x^2$, and $x^3$ terms of the intra-pixel sensitivity show very strong correlations (>0.9 or < -0.9) amongst themselves and with the star flux. Removing any of these parameters results in a larger BIC value.

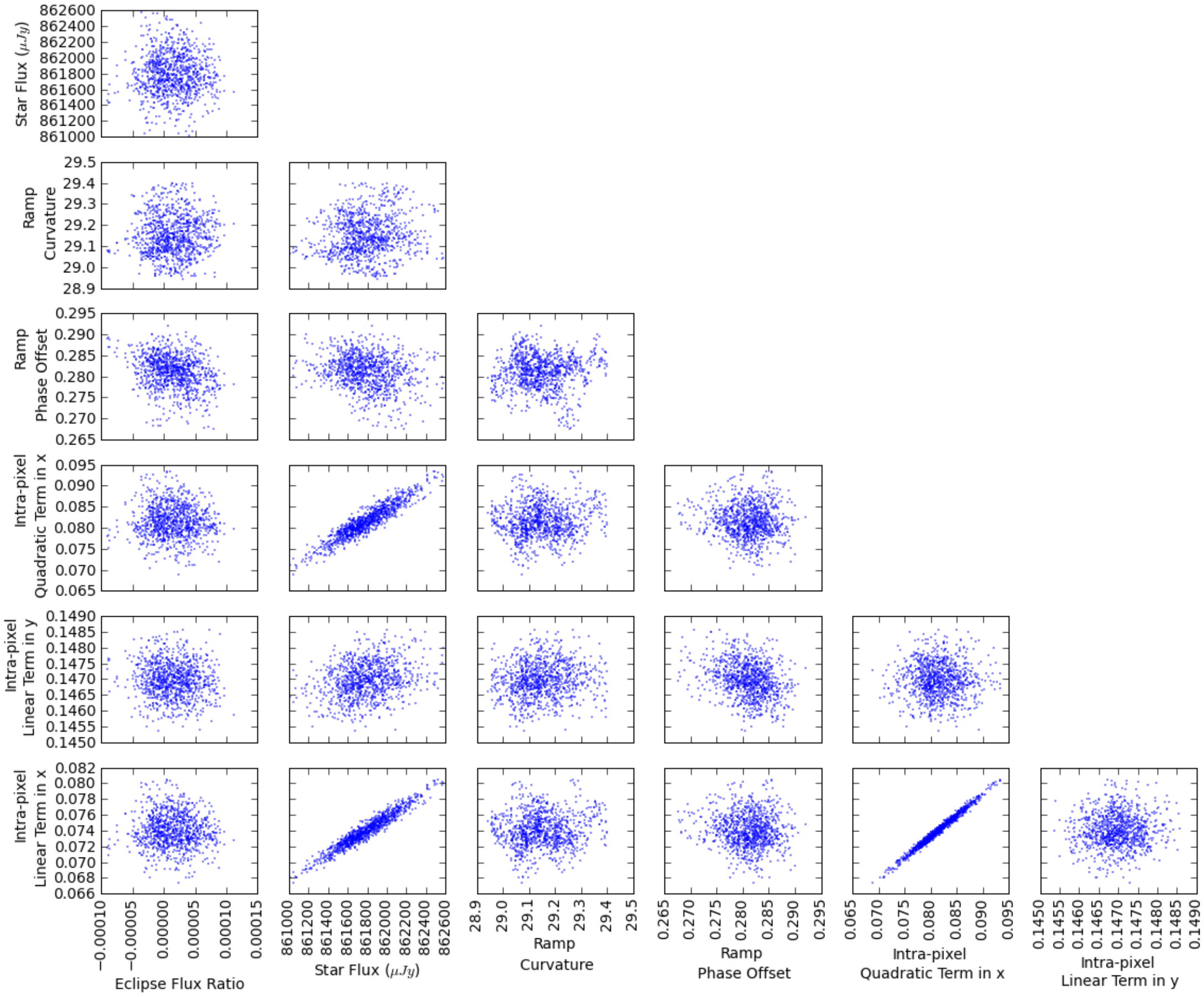

**Supplementary Figure 13. Phase-space projections at every 1000th MCMC step at 4.5 μm.** There are correlations of 0.96, 0.93, and 0.99 between the star flux and the *x* term of the intra-pixel sensitivity, the star flux and the $x^2$ term, and the $x^2$ and *x* terms, respectively. Again, removing one or more of these parameters results in a larger BIC value.

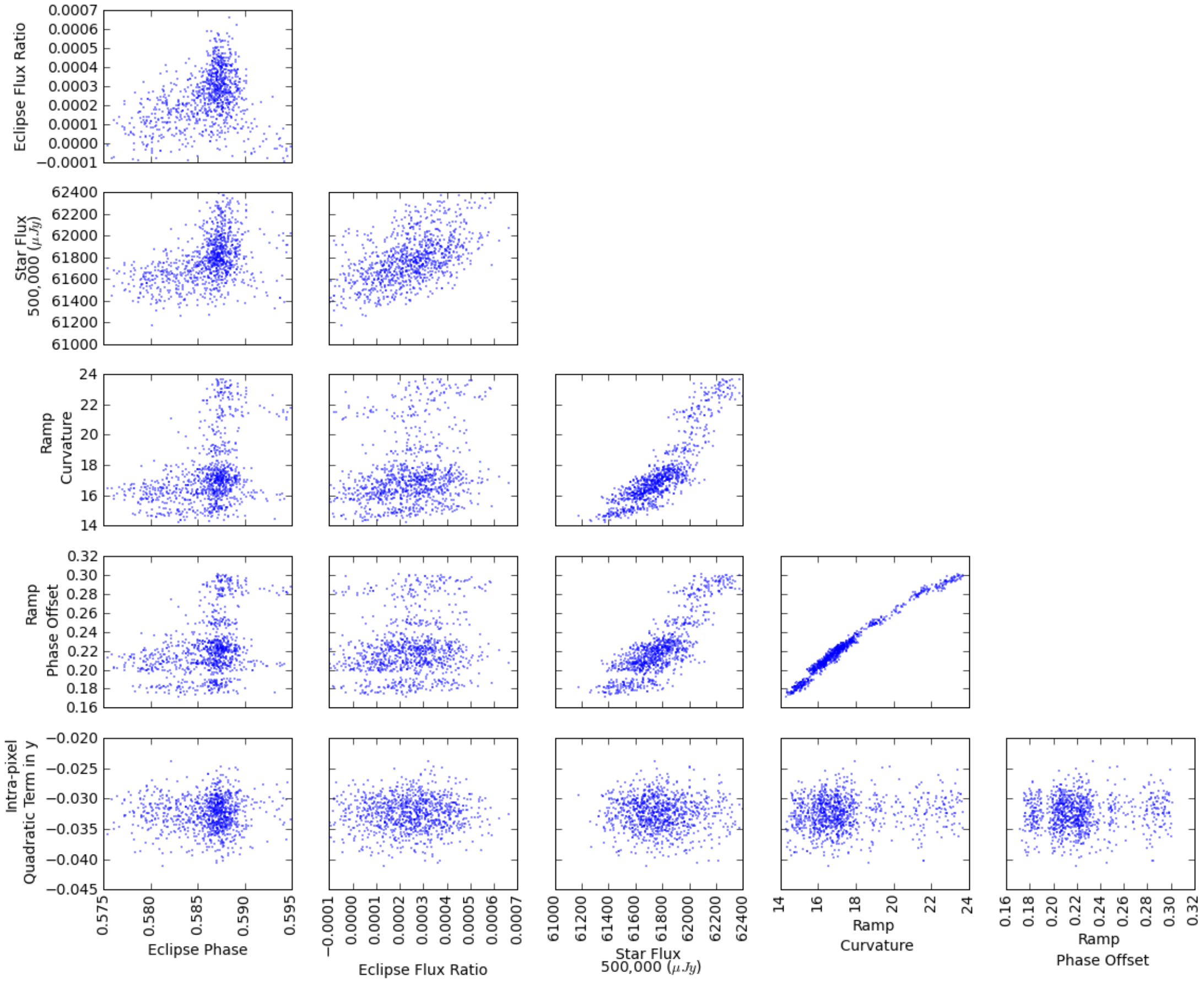

**Supplementary Figure 14. Phase-space projections at every 1000$^{th}$ MCMC step at 5.8 μm.** The ramp curvature and phase offset show a correlation of 0.97. Neither parameter can be removed without deteriorating the fit.

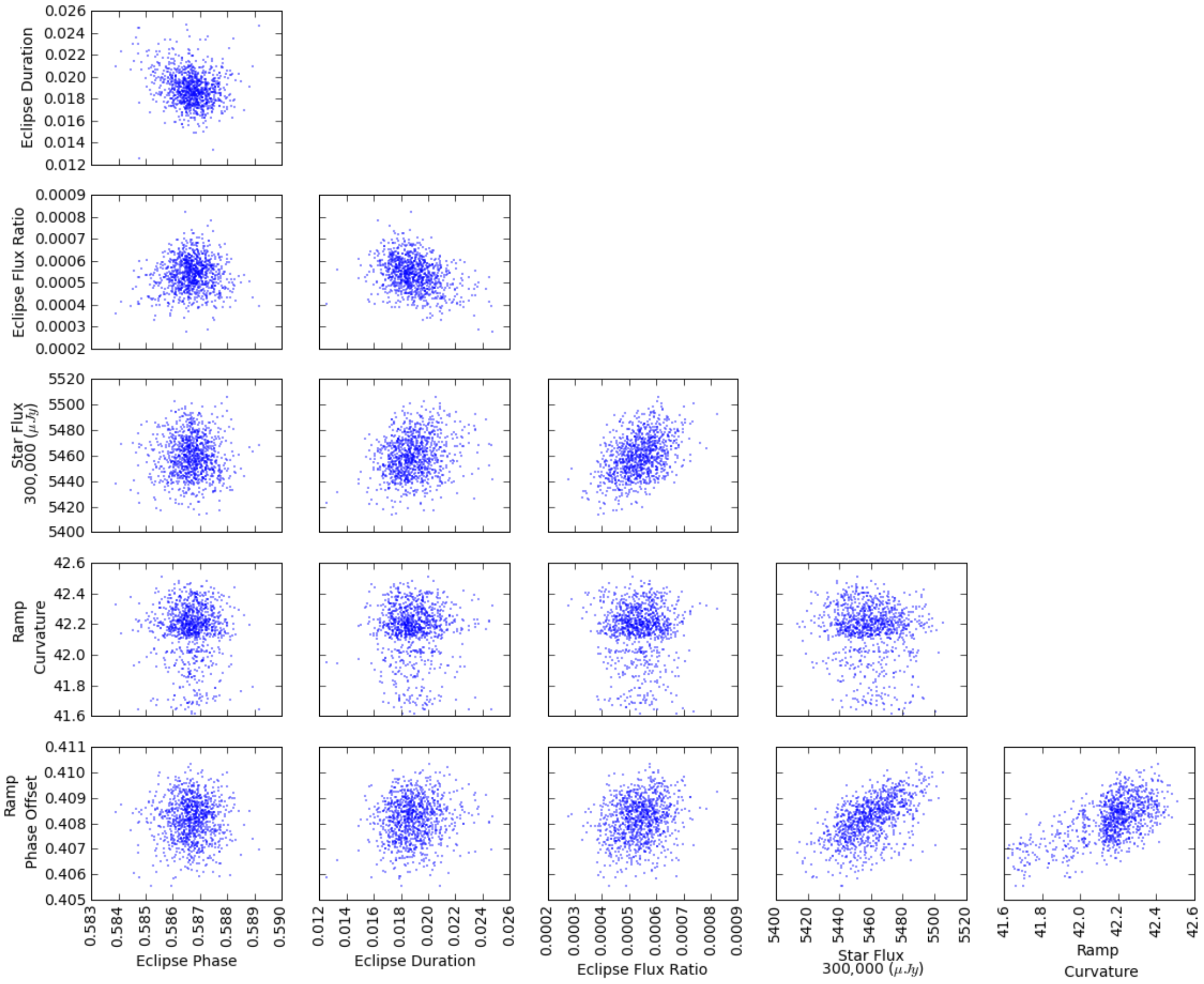

**Supplementary Figure 15. Phase-space projections at every 1000th MCMC step at 8.0 μm.**

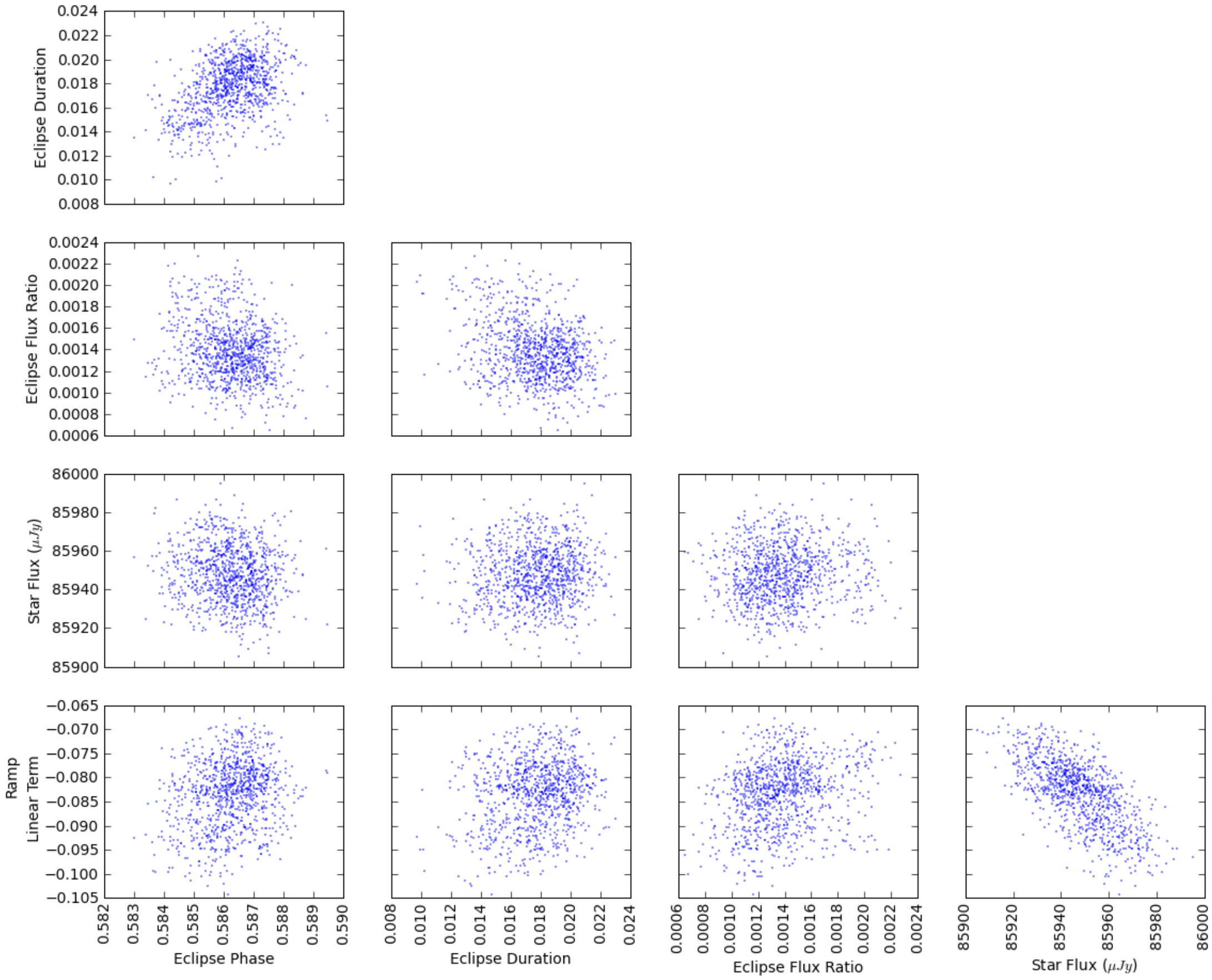

**Supplementary Figure 16. Phase-space projections at every 100th MCMC step at 16 μm.**

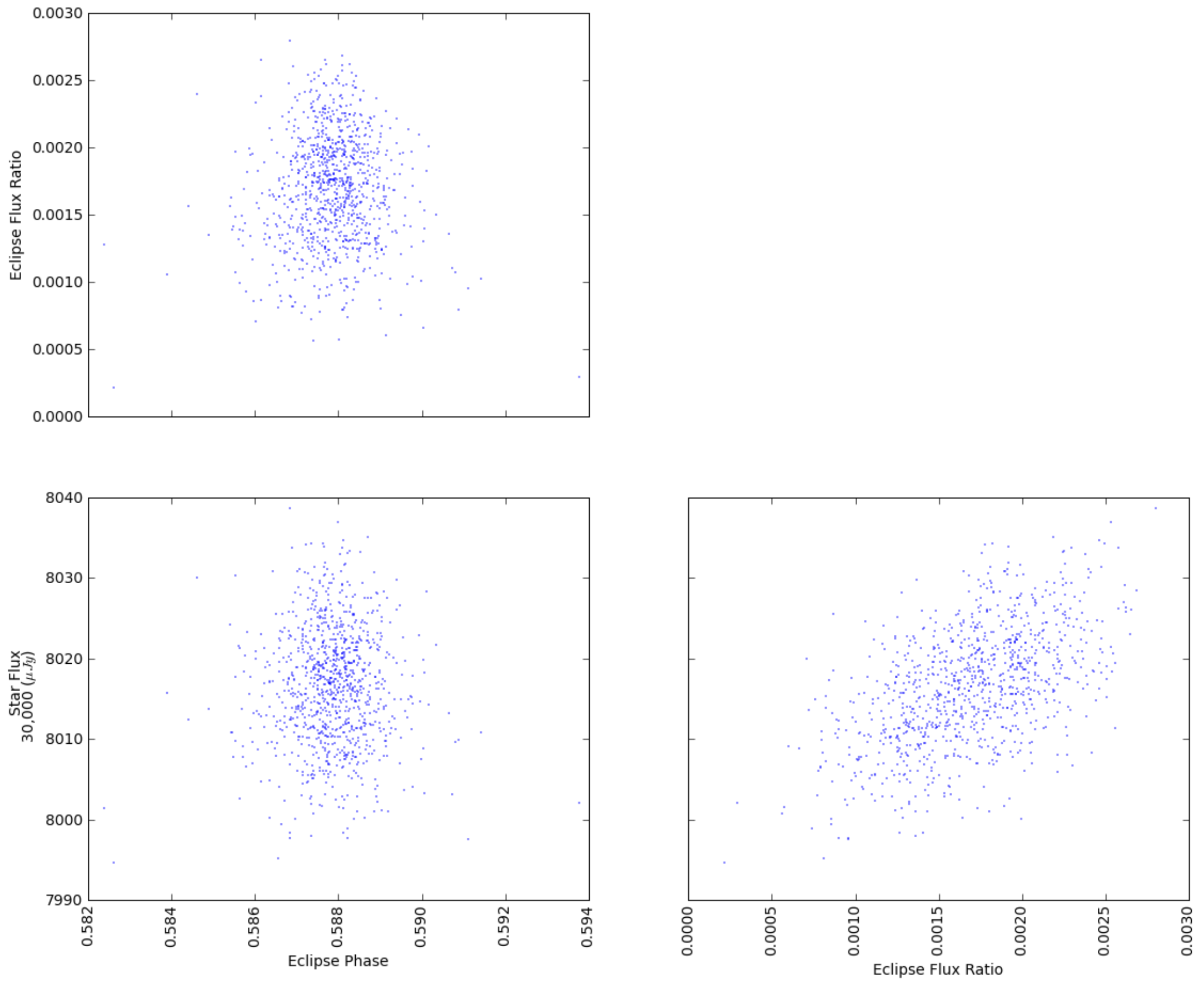

**Supplementary Figure 17. Phase-space projections at every 100th MCMC step at 24 μm.**

## Supplementary References: